\documentclass[fleqn,usenatbib]{mnras}

\usepackage[T1]{fontenc}
\usepackage[utf8]{inputenc}
\usepackage{xcolor}

\DeclareRobustCommand{\VAN}[3]{#2}
\let\VANthebibliography\thebibliography
\def\thebibliography{\DeclareRobustCommand{\VAN}[3]{##3}\VANthebibliography}

\usepackage{acronym}
\usepackage{booktabs}
\usepackage{dcolumn}
\usepackage{makecell}
\usepackage{graphicx}
\usepackage{tabularx}
\usepackage{amsmath,amsfonts,amssymb}
\usepackage{mathrsfs}
\usepackage[normalem]{ulem}
\usepackage{multirow}
\usepackage{color}
\usepackage{natbib}
\usepackage{verbatim}
\usepackage{physics}
\usepackage{braket}
\usepackage{colortbl}
\usepackage{newtxtext,newtxmath}

\DeclareOldFontCommand{\rm}{\normalfont\rmfamily}{\mathrm}
\DeclareOldFontCommand{\sf}{\normalfont\sffamily}{\mathsf}
\DeclareOldFontCommand{\tt}{\normalfont\ttfamily}{\mathtt}
\DeclareOldFontCommand{\bf}{\normalfont\bfseries}{\mathbf}
\DeclareOldFontCommand{\it}{\normalfont\itshape}{\mathit}
\DeclareOldFontCommand{\sl}{\normalfont\slshape}{\@nomath\sl}
\DeclareOldFontCommand{\sc}{\normalfont\scshape}{\@nomath\sc}
\DeclareRobustCommand*\cal{\@fontswitch\relax\mathcal}
\DeclareRobustCommand*\mit{\@fontswitch\relax\mathnormal}

\AtBeginDocument{\mathcode`v=\varv}

\definecolor{cyan}{rgb}{0,0.9,0.9}
\definecolor{orange}{rgb}{0.9,0.5,0}
\definecolor{magenta}{rgb}{1,0,1}
\definecolor{purple}{rgb}{0.8,0.4,0.8}
\definecolor{darkgreen}{rgb}{0.0,0.5,0.0}
\definecolor{gray}{rgb}{0.8242,0.8242,0.8242}
\definecolor{cadmiumgreen}{rgb}{0.0, 0.42, 0.24}
\definecolor{olive}{rgb}{0.5, 0.5, 0.0}

\newcommand{\reftab}[1]{{Table~\ref{#1}}}
\newcommand{\refsec}[1]{{Sec.~\ref{#1}}}
\newcommand{\reffig}[1]{{Fig.~\ref{#1}}}

\newcommand{\refeq}[1]{{Eq.~(\ref{#1})}}

\title[xkn: a semi-analytic kilonova framework]{xkn: a semi-analytic framework for the modelling of kilonovae}

\author[G.~Ricigliano et al.]{
Giacomo Ricigliano$^{1}$\thanks{Contact e-mail: giacomo.ricigliano@gmail.com},
Albino Perego$^{2,3}$,
Ssohrab Borhanian$^{4}$,
Eleonora Loffredo$^{5,6}$, \newauthor
Kyohei Kawaguchi$^{7,8}$,
Sebastiano Bernuzzi$^{4}$,
Lukas Chris Lippold$^{4}$
\\
$^{1}$ Institut für Kernphysik, Technische Universität Darmstadt, Schlossgartenstr. 2, Darmstadt 64289, Germany\\
$^{2}$ Dipartimento di Fisica, Universit\'{a} di Trento, Via Sommarive 14, 38123 Trento, Italy \\
$^{3}$ INFN-TIFPA,Trento Institute for Fundamental Physics and Applications, via Sommarive 14, I-38123 Trento, Italy \\
$^{4}$ Theoretisch-Physikalisches Institut, Friedrich-Schiller Universit{\"a}t Jena, 07743, Jena, Germany\\
$^{5}$ Gran Sasso Science Institute, Viale Francesco Crispi 7, 67100 L'Aquila, Italy \\
$^{6}$ INFN - Laboratori Nazionali del Gran Sasso, Via G. Acitelli 22, 67100 Assergi L'Aquila, Italy \\
$^{7}$Max Planck Institute for Gravitational Physics (Albert Einstein Institute), Am M\"{u}hlenberg 1, Potsdam-Golm, 14476, Germany\\
$^{8}$Institute for Cosmic Ray Research, The University of Tokyo, 5-1-5 Kashiwanoha, Kashiwa, Chiba 277-8582, Japan\\
}

\date{Accepted XXX. Received YYY; in original form ZZZ}

\pubyear{2022}

\begin{document}
\label{firstpage}
\pagerange{\pageref{firstpage}--\pageref{lastpage}}
\maketitle

\begin{abstract}
After GW170817, kilonovae have become of great interest for the astronomical, astrophysics and nuclear physics communities, due to their potential in revealing key information on the compact binary merger from which they emerge, such as the fate of the central remnant or the composition of the expelled material.
Therefore, the landscape of models employed for their analysis is rapidly evolving, with multiple approaches being used for different purposes.
In this paper, we present \texttt{xkn}, a semi-analytic framework which predicts and interprets the bolometric luminosity and the broadband light curves of such transients.
\texttt{xkn} models the merger ejecta structure accounting for different ejecta components and non-spherical geometries.
In addition to light curve models from the literature based on time scale and random-walk arguments, it implements a new model, \texttt{xkn-diff}, which is grounded on a solution of the radiative transfer equation for homologously expanding material.
In order to characterize the variety of the ejecta conditions, it employs time and composition dependent heating rates, thermalization efficiencies and opacities.
We compare \texttt{xkn} light curves with reference radiative transfer calculations, and we find that \texttt{xkn-diff} significantly improves over previous semi-analytic prescriptions.
We view \texttt{xkn} as an ideal tool for extensive parameter estimation data analysis applications.
\end{abstract}

\begin{keywords}
	stars: neutron -- methods: analytical, numerical 
\end{keywords}

\section{Introduction}

The detection of electromagnetic counterparts of gravitational wave signals represents one of the key aspects of gravitational wave astrophysics and, more in general, of multimessenger astronomy. While the gravitational wave signal produced by a coalescing compact binary encodes many properties and information about the merging system (e.g. the chirp mass, the masses of the two compact objects or their tidal deformation, if at least one of the two is not a black hole),
the electromagnetic signal can provide complementary information, including for example the amount of matter expelled during the merger and its chemical composition.
Other aspects of the merger, such as the nature of the coalescing objects or of the remnant that forms after the merger, could affect both the gravitational and the electromagnetic emission. In this case, the presence of more than one signal can provide tighter constraints and help discriminating between ambiguous or degenerate situations \citep[see e.g.][]{Radice:2018ozg,Hinderer:2018pei,Barbieri:2019bdq,Barbieri:2020ebt,Dietrich:2020efo,Raaijmakers:2021slr}.

The potential of multimessenger astrophysics was recently revealed by GW170817 \citep{TheLIGOScientific:2017qsa,GBM:2017lvd,Abbott:2018wiz}.
Just from the analysis of the gravitational wave signal, it was impossible to exclude that the coalescing objects were two black holes, since the posterior of the binary tidal deformability was extending down to 0 \citep{TheLIGOScientific:2017qsa,Abbott:2018wiz}, the value expected for a binary black hole. The identification of the system as a binary neutron star was mostly based on the values of the masses of the merging bodies and, more importantly, on the detection of two electromagnetic counterparts, namely a short gamma-ray burst and a kilonova 
\citep{Chornock:2017sdf,Cowperthwaite:2017dyu,Coulter:2017wya,Drout:2017ijr, Evans:2017mmy, Goldstein:2017mmi, Hallinan:2017woc, Kasliwal:2017ngb, Murguia-Berthier:2017kkn, Nicholl:2017ahq, Smartt:2017fuw, Soares-Santos:2017lru, Tanvir:2017pws, Troja:2017nqp, Villar:2017wcc, Waxman:2017sqv,Ghirlanda:2018uyx}.
Such a combination of signals was indeed very useful in providing constraints on the equation of state of nuclear matter or in shedding light on the central engine of gamma-ray bursts \citep[see e.g.][]{De:2018uhw,Abbott:2018exr,Coughlin:2018fis}.
On the other hand, in the case of the subsequent binary neutron star merger, GW190425 \citep{Abbott:2020uma}, or in the first observed black hole-neutron star systems \citep{LIGOScientific:2021djp,LIGOScientific:2021qlt}, no electromagnetic counterparts were observed \citep[see e.g.][]{Coughlin:2019zqi}. 
In these cases, the nature of the binary was deduced only from the inferred masses, while the gravitational signal alone was not informative on the tidal deformability of the system, due to the lower expected values and to the not sufficiently high signal-to-noise ratio. 

Among the different electromagnetic counterparts, the kilonova is one of the most peculiar transients associated to compact binary mergers involving at least one neutron star \citep{Li:1998bw,Metzger:2010sy}. It arises when the merger and its remnant expel a non-negligible amount of neutron-rich matter, which undergoes $r$-process nucleosynthesis \citep{Lattimer:1974slx,Eichler:1989ve,Freiburghaus:1999}; see also \citet{Cowan:2019pkx,Perego:2021dpw} for recent reviews. The decay of the freshly produced radioactive elements moving from the neutron-rich side towards the bottom of the nuclear valley of stability releases nuclear energy that, despite the fast expansion, keeps the expanding ejecta hot. Due to expansion the matter opacity to the electromagnetic radiation decreases until photons can eventually diffuse out, producing a kilonova \citep[see][for a recent review and references therein]{Metzger:2019zeh}.
Depending on the mass of the ejecta, on their expansion velocity and  composition, the peak of the kilonova emission is expected to occur between a few hours and several days after the merger \citep[see, for example,][]{Arnett:1982,Pinto_2000,Barnes:2013wka,Tanaka:2013ana,Radice:2018pdn,Kawaguchi:2019nju}. At the same time, the emission is expected to evolve, moving from bluer to redder frequencies as a consequence of the photospheric expansion, of the decrease in the nuclear energy input and in the opacity of matter, as well as of the viewing angle \citep[see e.g.][]{Metzger:2014ila,Perego:2014qda,Martin:2015hxa,Rosswog:2016dhy,Wollaeger:2017ahm,Kasen:2018drm,Korobkin:2020spe}. As long as the opacity of the innermost ejecta is large enough, the kilonova is characterized by the presence of a photosphere and the resulting emission can be described, in good approximation, as quasi-thermal. Non-thermal and non-local thermodynamics equilibrium effects become more and more relevant as time increases, until the transient enters its nebular phase. In the case of AT2017gfo (the kilonova associated to GW170817) the transition from a full photospheric regime to the nebular phase happened between a few days to a week after merger \citep[see, for example,][]{Smartt:2017fuw,Watson:2019xjv,Wu:2018mvg,Pognan:2021wpy,Pognan:2022pix,Gillanders:2023jpd}.

The modelling of kilonovae is extremely challenging. It requires the solution of a radiative transfer (RT) problem in a fast expanding, radioactive and radiation dominated medium. Not only the composition of matter changes with time due to nuclear reactions and decays, but due to the expansion and to the interaction between matter and radiation, atoms inside the ejecta (which are initially fully ionized due to the large matter temperature) span different ionization levels, following the progressive electron recombination. The presence of heavy elements, and in particular, of lanthanides and actinides, largely increase the photon opacity due to bound-bound and bound-free transitions involving the $f$ electron shells \citep{Roberts:2011xz,Kasen:2013xka,Tanaka:2013ana}. For most of the heavy elements, opacities due to ionized species and for matter in the thermodynamics regime relevant for kilonovae are experimentally unknown and their values are usually provided by non-trivial atomic structure calculations \citep[see, for example,][]{Fontes:2019tlk,Tanaka:2019iqp,Banerjee:2022doa,Banerjee:2023gye}. Furthermore, large uncertainties still affect the calculation of the detailed nuclear energy released by $r$-process elements \citep{Rosswog:2017sdn,Barnes:2020nfi,Zhu:2020eyk,Zhu:2022qhc,Lund:2022bsr}, as well as the estimation of the fraction of the energy that the expanding matter is able to thermalize \citep{Hotokezaka:2015cma,Barnes:2016umi,Kasen:2018drm}.
In addition to the difficulties related with the problem of transporting photons inside an expanding, radioactive medium, an additional challenge is represented by the fact that a binary neutron star merger or a black hole-neutron star merger can expel matter with different properties and, possibly, with a high degree on anisotropy \citep[see, e.g., ][]{Wanajo:2014wha,Just:2014fka,Sekiguchi:2016bjd,Foucart:2016rxm,Perego:2017wtu}. This implies that the medium inside which the photons are produced, diffused and emitted can have a non-trivial stratification, as well as angular distribution.

It is not surprising that, given the complexity of the kilonova scenario and the variety of aspects involved, so far the problem of predicting or producing kilonova light curves and spectra has been tackled by a large variety of models, employing very different levels of sophistications and approximations. 
Some models solve the photon transport problem in an expanding medium considering wavelenght- and composition-dependent opacities, computed consistently and coupled to the calculation of the abundances of the different ion species, assuming local thermodynamics equilibrium \citep[see, e.g., ][]{Tanaka:2013ana,Kasen:2017sxr,Wollaeger:2017ahm,Shingles:2023kua}.
These models are the most sophisticated and reliable ones, but they necessarily require large computational resources, especially in three dimensions. 
Other examples of kilonova models include TARDIS \citep{Kerzendorf:2014}, which solves the 1D photon transport problem in the optically thin atmosphere above a predefined photosphere, POSSIS \citep{Bulla:2019muo,Bulla:2022mwo}, which uses pre-computed wavelenght- and time-dependent opacities on a 3D Cartesian grid,
or SNEC \citep{Morozova:2015bla,Wu:2021ibi}, which solves radiation hydrodynamics in spherical symmetry through a gray flux-limited diffusion approach.
These more approximated approaches clearly reduce the computational effort, especially if some symmetry is invoked.

At the opposite extreme, kilonova light curves have also been computed by using simplified kilonova models that avoid the direct solution of the RT problem, since they are often based on the solution of the energy conservation equation inside the ejecta or on time scale arguments mimicking the mean features of the photon diffusion problem \citep[see, e.g., ][]{Grossman:2013lqa,Martin:2015hxa,Hotokezaka_2020}. They usually employ gray, constant opacities and can reproduce some of the most relevant features of the kilonova emission, at least at a qualitative level. The extremely reduced computational costs of these models allows their usage in multi-dimensional parameter estimate analysis, which requires the evaluation of millions, if not billions of kilonova light curves \citep[as done, e.g., by ][]{Breschi:2021tbm}.

With the increase of the number and sensitivity of gravitational wave detectors, the amount of multimessenger signals, and in particular, of kilonova counterparts of gravitational wave events, is expected to significantly grow in the years to come. For example, during the fourth observational campaign of LIGO, Virgo and KAGRA, the number of detected binary neutron star merger is expected to be a few tens per year \citep{Petrov:2022,Colombo:2022zzp}. 
Additionally, the careful (re)analysis of the afterglow signals of close-by gamma-ray bursts can reveal signatures of kilonova emission, as in the case of the exceptionally bright GRB211211A \citep{Troja:2022yya}, or more recently also in the case of the long GRB230307A \citep{Levan:2023ssd}.
Given the present scenario, characterized by a growing number of kilonovae, which could significantly differ in terms of intrinsic properties, as well as in the quality and quantity of the data, it is still imperative to improve on the accuracy of fast and approximated kilonova models. The latter can be complementary to more sophisticated models, since they can be used for extensive parameter estimations, and to provide a robust and reliable framework to analyze coherently kilonova emissions coming from very different events. They can also be coupled to gravitational wave data analysis in the quest for coherent and genuine multimessenger analysis.

In this paper, we present \texttt{xkn}, a semi-analytic framework to perform analysis of kilonova emission, both in terms of bolometric luminosity and broadband light curves. \texttt{xkn} inherits the possibility of adding several ejecta components and of prescribing non-trivial ejecta geometries from previous implementations \citep{Martin:2015hxa,Perego:2017wtu}. However, with respect to the latter, it aims to improve on the accuracy and on the reliability of the resulting light curves by replacing the kilonova model grounded on time scale arguments with a different model, \texttt{xkn-diff}, based on a semi-analytic solution of the diffusion equation for homologously expanding ejecta.
Such a solution was presented in \citet{Wollaeger:2017ahm}, and based on works reported in \citet{Pinto_2000}. Here, we expand the class of solutions, by considering time dependent heating rates, thermalization efficiencies and opacities. Moreover, we improve the physical input, by including composition dependent heating rates and opacities.

The paper is structured as follows: in \refsec{sec:1d_model} we present in detail the spherically symmetric, kilonova emission model \texttt{xkn-diff}, distinguishing between the optically thick (\refsec{subsec:op_thick}) and the optically thin (\refsec{subsec:op_thin}) part. The general multi-component, anisotropic framework of \texttt{xkn} to compute light curves is presented in \refsec{sec:xkn framework}, while in \refsec{sec:input_physics} we detail the input physics entering the model, listing the heating rates (\refsec{subsec:heating_rates}) and the opacity (\refsec{subsec:opacities}) prescriptions. In \refsec{sec:RT_comparison} we compare the results of the various \texttt{xkn} models with the ones obtained using a RT code \citep{Tanaka:2013ana,Kawaguchi:2018ptg,Kawaguchi:2020vbf}, taken as reference, in order to address the degree of accuracy and the limitations of our approach. We provide a summary and the conclusions of our work in \refsec{sec:conclusions}.

\section{Semi-analytic 1D kilonova model}\label{sec:1d_model}
Our kilonova model is based on a one-dimensional model for the diffusion and emission of photons from homologously expanding, radioactive matter. More specifically, the kilonova emission is calculated as the combination of two contributions, one emitted at the ejecta photosphere, i.e. the surface delimiting the optically thick bulk of the ejecta and from which photons can escape and move inside the atmosphere, and one coming from the optically thin layers above it. In the following, we separately present these two contributions.

\subsection{Optically thick ejecta treatment} \label{subsec:op_thick}
The contribution to the luminosity arising from the photosphere is computed starting from the semi-analytic formula originally proposed by \citet{Pinto_2000} with the intent to treat the ejecta from Type Ia supernovae (SNe Ia), and later adapted by \citet{Wollaeger:2017ahm} to model kilonovae. In spite of its simplifying assumptions, this formula can qualitatively reproduce the thermal evolution of the ejecta. In the following, we report in broad lines its derivation.\\

The ejecta fluid is assumed to be optically thick throughout its entire depth and the radiation field properties are evolved on the basis of the time-dependent equation of RT.
In particular, we consider the first two frequency-integrated moments of such equation in the comoving frame, calculated to order $O(v/c)$:
\begin{equation}\label{eq:rad_trans_E}
\begin{split}
    & \frac{DE}{Dt}+\frac{1}{r^2}\frac{\partial}{\partial r}(r^2F)+\frac{v}{r}(3E-P)+\frac{\partial v}{\partial r}(E+P)= \\
    & =\int_0^{\infty}(4\pi\eta_{\nu}-c\chi_{\nu}E_{\nu})d\nu \, ,
\end{split}
\end{equation}
\begin{equation}\label{eq:rad_trans_F}
    \frac{1}{c^2}\frac{DF}{Dt}+\frac{\partial P}{\partial r}+\frac{3P-E}{r}+\frac{2}{c^2}\left(\frac{\partial v}{\partial r}+\frac{v}{r}\right)F=-\frac{1}{c}\int_0^{\infty}\chi_{\nu}F_{\nu}d\nu \, ,
\end{equation}
where $v$ is the fluid velocity field, $r$ the radial coordinate, $\chi_{\nu}$ the extinction coefficient and $\eta_{\nu}$ the volume emissivity, while the subscript $\nu$ indicates the frequency dependence. Here $E$, $F$ and $P$ are the energy density, flux and radiation pressure of the radiation field, respectively. $D/Dt$ indicates the comoving (Lagrangian) derivative.\\
The solution of this set of equations is found by adopting a series of hypotheses. 
\begin{enumerate}
\item We assume a homologous expansion of the fluid, i.e. each fluid element expands with constant radial speed. Under this assumption, the ejecta maintain their proportions while expanding with an external radius $R(t)\simeq v_{\mathrm{max}}t$, where $v_{\mathrm{max}}$ is the maximum outflow velocity. The homologous expansion hypothesis is consistent as long as the energy heating the fluid does not affect the fluid motion in a significant way.
\item We resort to the Eddington approximation, wherein the radiation field is isotropic and the relation $E=3P$ holds. The latter is valid since the outflow is optically thick, as one can expect at least in the early stages of its evolution. When the fluid becomes transparent at later times, even if this approximation breaks down, the error has still little effect on the bolometric luminosity.
\item Regarding the energy balance, we assume that the gas internal energy, that is its thermal kinetic energy as well as the ionization energy, is subdominant with respect to the radiation field energy, and therefore we ignore the former (radiation dominated conditions).
\item Additionally, we assume that the absorbed heat is immediately re-radiated as thermal emission, and thus:
\begin{equation}
    \int_0^{\infty}(4\pi\eta_{\nu}-c\chi_{\nu}E_{\nu})d\nu=\dot{E}_{\rm heat} \, ,
\end{equation}
where $\dot{E}_{\rm heat}$ is the energy deposition rate per unit volume.
\end{enumerate}
In light of the above considerations, we act on \refeq{eq:rad_trans_E} and \refeq{eq:rad_trans_F} with the aim to simplify them. \refeq{eq:rad_trans_E} is an equation for the energy density field, $E(r,t)$, and in order to ensure the correct radiation energy balance we retain all terms to the order $O(v/c)$, as $E(r,t)$ changes considerably on the hydrodynamic timescale. \refeq{eq:rad_trans_F} is instead an equation for the radiation momentum $F(r,t)$ and we solve it at lower order by discarding all time and velocity-dependent terms. This choice is appropriate on the fluid-flow timescale as we assumed that $F(r,t)$ is not relevant for the outflow dynamics. Hence by inverting \refeq{eq:rad_trans_F} we obtain:
\begin{equation}\label{eq:flux}
    F=-\frac{c}{3\chi}\frac{\partial E}{\partial r} \, ,
\end{equation}
with $\chi$ a properly defined frequency-averaged inverse mean free path. The above expression for the flux can be inserted in \refeq{eq:rad_trans_E}, resulting in:
\begin{equation}\label{eq:diffusion}
    \frac{DE}{Dt}-\frac{c}{3r^2}\frac{\partial}{\partial r}\left(\frac{r^2}{\chi}\frac{\partial E}{\partial r}\right)+\frac{4\dot{R}}{R}E=\dot{E}_{\rm heat} \, ,
\end{equation}
where $\dot{R}$ denotes the derivative with respect to time.
We now introduce $\kappa$ as an absorption opacity, homogeneous in space (but not necessarily in time), such that $\chi=\kappa\rho$, with $\rho$ the fluid density.
Moreover, we express $\dot{E}_{\rm heat}$ as $\dot{E}_{\rm heat} = \dot{\epsilon} f_{\rm th}$, where 
$\dot{\epsilon}$ is the energy release rate per unit mass and $f_{\rm th}$ a thermalization efficiency coefficient, and both are function of time.
From the assumption of homologous expansion, we recall that:
\begin{equation}\label{eq:homologous}
    v=\frac{r}{t} \, , \quad v=v_{\mathrm{max}}x \, ,
\end{equation}
where $x \in [0,1]$ is the dimensionless radius coordinate. Moreover, in the radiation transport equation we adopt the following single-zone homologous solution for the expansion profile:
\begin{equation}\label{eq:density}
    \rho(t)=\rho_0\left(\frac{t_0}{t}\right)^3 \, ,
\end{equation}
with $t_0$ the initial time of the expansion, and $\rho_0$ the density at $t_0$. We approximate the latter as:
\begin{equation}
    \rho_0=\frac{M_{\mathrm{ej}}}{\frac{4}{3}\pi(v_{\mathrm{max}}t_0)^3} \, ,
\end{equation}
where $M_{\mathrm{ej}}$ is the ejecta mass.\\
Furthermore, from the hypothesis of radiation dominated gas, we employ the polytropic equation of state in the Eddington approximation to obtain $E\propto t^{-4}$. If we assume that the residual dependences of $E$ can be separated into a spatial profile $\psi(x)$ and a temporal profile $\phi(t)$, we can write:
\begin{equation}\label{eq:energy_density}
    E(x,t)=E_0\left(\frac{t_0}{t}\right)^4\psi(x)\phi(t) \, ,
\end{equation}
where $E_0$ is treated as a free parameter. In particular, assuming that the radiation field has a black-body spectrum, we relate it to an initial black-body temperature:
\begin{equation}
    T_0=\left(\frac{E_0}{a}\right)^{\frac{1}{4}} \, ,
\end{equation}
with $a=4\sigma_{\mathrm{SB}}/c=7.5657\times10^{-15}$ erg $\mathrm{cm^{-3}K^{-4}}$ being the radiation constant and $\sigma_{\mathrm{SB}}$ the Stefan-Boltzmann constant.
Using \refeq{eq:homologous}, \refeq{eq:density} and \refeq{eq:energy_density}, the transport equation \refeq{eq:diffusion} becomes:
\begin{equation}\label{eq:diffusion2}
    \left( \frac{E_0t_0}{\rho_0} \right) \frac{1}{t} \left[ \psi(x)\phi'(t)- \frac{ t}{t_0 \tau } \phi(t) \frac{1}{x^2}\left(x^2\psi'(x)\right)' \right] = f_{\rm th} \dot{\epsilon}_r \, ,
\end{equation}
with the prime superscript on a function indicating the derivative with respect to its variable and where
\begin{equation}
    \tau \equiv \frac{3\kappa\rho_0}{c}(v_{\mathrm{max}}t_0)^2 \, .
\end{equation}
The latter is a comprehensive factor comparable to the diffusion time scale that carries a possible dependence on $t$ through $\kappa$.
The homogeneous form of \refeq{eq:diffusion2} can be solved by means of variable separation, according to which the resulting two functions in $x$ and $t$ must be identically equal to the same separation constant, $\lambda$:
\begin{equation}\label{eq:diffusion_split}
    \frac{1}{x^2\psi(x)}\left(x^2\psi'(x)\right)'=-\lambda \, , \qquad\tau_0\left(\frac{t_0}{t}\right)\frac{\phi'(t)}{\phi(t)}=-\lambda \, .
\end{equation}
resulting in two ordinary differential equations to be solved.
The equation for the spatial profile can be expressed as an eigenvalue equation for the operator $A$:
\begin{equation}\label{eq:eigeneq}
    A\psi(x) \equiv -\frac{1}{x^2}\left(x^2\psi'(x)\right)'=\lambda\psi(x) \, .
\end{equation}
The eigenfunctions of \refeq{eq:eigeneq} can be determined by imposing suitable boundary conditions to the problem. While for the temporal profile of the energy density $E(x,t)$ we naturally assume $\phi(t_0)=1$ and $\phi(\infty)=0$, for the spatial part it is reasonable to consider a reflection symmetry at $x=0$ and a radiative-zero condition at $x=1$, being the outflow optically thick:
\begin{equation}\label{eq:conditions}
    \psi'(0)=0 \, , \qquad\psi(1)=0 \, .
\end{equation}
\refeq{eq:conditions} can be directly translated into identical conditions for the eigenfunctions of \refeq{eq:eigeneq}. If we impose the normalization requirement:
\begin{equation}
    \braket{\psi_n|\psi_m}=\delta_{n,m} \, ,
\end{equation}
as we adopt the notation:
\begin{equation}
    \braket{f|g}=\int_0^1f(x)g(x)x^2dx \, ,
\end{equation}
the resulting homogeneous spatial eigenfunctions are:
\begin{equation}\label{eq:eigensol}
    \psi_n(x)=\sqrt{2}~\frac{\sin(n\pi x)}{x} \, ,
\end{equation}
with $\lambda=n^2\pi^2$ and $n$ any positive integer.\\
\refeq{eq:eigensol} represents a complete orthonormal basis on the interval of interest, and therefore we can expand the general solution of \refeq{eq:diffusion2} on the latter:
\begin{equation}
    E(x,t)=\sum_{n=1}^{\infty}c_n(t)\psi_n(x) \, .
\end{equation}
Here the coefficients $c_n(t)$ retain a generic time dependence, and we can conveniently redefine them in the form:
\begin{equation}
    c_n(t)=E_0\left(\frac{t_0}{t}\right)^4\phi_n(t) \, ,
\end{equation}
thus obtaining:
\begin{equation}\label{eq:expansion}
    E(x,t)=E_0\left(\frac{t_0}{t}\right)^4\sum_{n=1}^{\infty}\phi_n(t)\psi_n(x) \, .    
\end{equation}
Using \refeq{eq:expansion} in \refeq{eq:diffusion2}, we can exploit the orthonormality of $\psi_n(x)$ integrating over $x\in[0,1]$ to find:
\begin{equation}\label{eq:temporal_eq}
   \phi_n'(t)+\left(\frac{t}{t_0\tau}\right)(n^2\pi^2)\phi_n(t) =\frac{(-1)^{n+1} \rho_0 \sqrt{2}}{n\pi E_0} 
     \left(\frac{t}{t_0}\right)  
   \dot{\epsilon} f_{\rm th} \, .
\end{equation}
Finally, the bolometric luminosity is found by employing \refeq{eq:flux} and \refeq{eq:expansion} to compute the flux at the surface of the ejecta:
\begin{equation}\label{eq:luminosity}
\begin{split}
    L(t) & =4\pi R^2(t)\left[x^2F(x,t)\right]_{x=1} \\
    & =\frac{4\pi \left(v_{\mathrm{max}}t_0 \right)^3\sqrt{2}E_0}{\tau}\sum_{n=1}^{\infty}(-1)^{n+1}n\pi\phi_n(t) \, ,
\end{split}
\end{equation}
where $\phi_n(t)$ are the solutions of \refeq{eq:temporal_eq},
that can be obtained once the time-dependence of $\dot{\epsilon}$, $f_{\rm th}$ and $\kappa$ has been specified.
If the formal solution of \refeq{eq:temporal_eq} is complex, when we compute the luminosity through \refeq{eq:luminosity} we take only the real part of it.

Since this model is valid in the limit of optically thick matter, the outcome of \refeq{eq:luminosity} is rescaled by a factor $M_{\mathrm{thick}}/M_{\mathrm{ej}}$, where $M_{\mathrm{thick}}$ is the mass of the optically thick portion of ejecta, defined as the region enclosed by the photosphere:
\begin{equation}
    M_{\rm thick}=4\pi\int_0^{R_{\rm ph}(t)}\rho(t,r)r^2dr \, .
\end{equation}
Differently from the single-zone approximation adopted in the solution of the RT equation, here  we choose a more accurate space-dependent density profile such as the self-similar homologous solution \citep{Wollaeger:2017ahm}:
\begin{equation}\label{eq:density_profile}
    \rho(t,x)=\rho_0\left(\frac{t_0}{t}\right)^3\left(1-x^2\right)^3 \, .
\end{equation}
The photospheric radius evolution $R_{\mathrm{ph}}(t)$ can be found analytically by imposing the condition
\begin{equation}\label{eq:photosphere_condition}
    \tau_{\gamma}(R_{\mathrm{ph}})=2/3 \, ,
\end{equation}
with $\tau_{\gamma}$ the optical depth of the material:
\begin{equation}
    \tau_{\gamma}(t,x)=\kappa\int_x^1\rho(t,x')dx' \, ,
\end{equation}
and $\rho(t,x)$ the density of \refeq{eq:density_profile}.
The determination of $R_{\rm ph}(t)$ implies the solution of a seventh order polynomial equation. However, regardless of the ejecta parameters, the temporal evolution of $R_{\rm ph}$ resembles a parabolic behaviour. Then, we approximate it as a parabolic arc with extremes fixed by the condition $R_{\mathrm{ph}}=0$ applied to \refeq{eq:photosphere_condition}, i.e.:
\begin{equation}
    t_1=0 \, , \qquad t_2=\sqrt{\frac{27M_{\rm ej}\kappa}{8\pi v_{\rm max}^2}} \, ,
\end{equation} 
and curvature fixed by a third point, $t_3$, taken in the proximity of $t_1$, where \refeq{eq:photosphere_condition} can be solved by assuming $\left(R_{\rm max}-R_{\mathrm{ph}} \right)/R_{\rm max} \ll 1$.
By adopting this approximate solution, the error on the photosphere position with respect to that provided by the exact solution of \refeq{eq:photosphere_condition} is contained within $8\%$.

We characterize the emission at the photosphere by assuming a Plankian black-body spectrum, and thus we compute the associated photospheric temperature $T_{\mathrm{ph}}(t)$ by means of the Stefan-Boltzmann law:
\begin{equation}\label{eq:photospheric_temperature}
    T_{\rm ph}(t)={\rm max}\left[\left(\frac{L_{\rm thick}(t)}{4\pi\sigma R_{\rm ph}^2(t)}\right)^{\frac{1}{4}},T_{\rm floor}\right] \, ,
\end{equation}
with $L_{\rm thick}(t)$ the luminosity of the thick part of the ejecta. A temperature floor $T_{\rm floor}$ is applied in order to approximately account for the electron-ion recombination during the ejecta expansion \citep[see e.g.][]{Barnes:2013wka,Villar:2017wcc}. When $T_{\mathrm{ph}}(t)$ reaches the temperature floor, $R_{\mathrm{ph}}(t)$ is thus recomputed solving the implicit equation obtained from the Stefan-Boltzmann law. The value of $T_{\rm floor}$, generally treated as model parameter, is in fact dependent on the ejecta opacity and therefore closely linked to its composition. For this reason, we also include the possibility to interpolate the floor temperature between two model parameters $T_{\rm Ni}$ and $T_{\rm La}$, corresponding to characteristic recombination temperatures in a Lanthanides-poor ($Y_e\gtrsim0.3$) and a Lanthanides-rich ($Y_e\lesssim0.2$) environment, respectively.

In the following, we discuss a few cases for which the solutions of \refeq{eq:temporal_eq} can be obtained analytically, based on temporal dependence of $\kappa$, $\epsilon$ and $f_{\rm th}$.

\subsubsection{Constant opacity, constant thermalization efficiency and power-law heating rate}
\label{subsec:heating power law}
We first consider the case in which $\kappa$ is not only uniform in space, but also constant in time, i,.e. $\kappa = \kappa_0$. In this case,
the quantity $\tau$ becomes a constant, $\tau_0 = 3 \kappa_0 \rho_0 (v_{\rm max}t_0)^2/c$. Additionally, we consider $f_{\rm th}$ to be a constant, $f_{{\rm th},0}$, while
\begin{equation}
\dot{\epsilon} = \dot{\epsilon}_0 \left( \frac{t}{t_0} \right)^{-\alpha} \, .
\label{eq:heating rate KN model}
\end{equation} 
In this case, the solutions of \refeq{eq:temporal_eq} take the explicit form:
\begin{equation}
    \phi_n(t)=\exp\left(-\frac{\pi^2n^2t^2}{2t_0\tau_0} \right)\left[K_n+A_n\Gamma\left(1-\frac{\alpha}{2},-\frac{\pi^2n^2t^2}{2t_0\tau_0}\right)\right] \, ,
    \label{eq:phi_n sol 1}
\end{equation}
where $\Gamma(s,x)$ is the upper incomplete gamma function, defined as:
\begin{equation}
    \Gamma(s,x) \equiv \int_x^{\infty}t^{s-1}e^{-t}dt \, ,
\end{equation}
while $A_n$ are constants:
\begin{equation}
    A_n= (-1)^{n+1+\frac{\alpha}{2}} \left(n \pi \right)^{\alpha-3} \sqrt{2}^{1-\alpha}
    \left( \frac{\rho_0 f_{\rm th,0} \dot{\epsilon}_0 \tau_0}{E_0} \right) \left( \frac{t_0}{\tau_0} \right)^{\frac{\alpha}{2}} \, ,   
    \label{eq:A_n sol 1} 
\end{equation}
and $K_n$ are integration constants fixed by the boundary conditions.\\
In order to find the latter, we exploit the assumptions over the temporal profile of the energy density. While $\phi(\infty)=0$ is automatically satisfied by the form of $\phi_n(t)$, one needs to translate $\phi(t_0)=1$ into a condition on $\phi_n(t)$. We choose to assign $\phi_n(t_0)=\delta_{n,1}$ to ensure the convergence of \refeq{eq:luminosity}:
\begin{equation}
K_n =  \delta_{n,1} \exp\left(-\frac{\pi^2 n^2 t_0}{2 \tau_0} \right) - 
A_n \Gamma\left(1-\frac{\alpha}{2},-\frac{\pi^2n^2t_0}{2 \tau_0}\right) \, .
\end{equation}

\subsubsection{Constant opacity, power-law thermalization efficiency and heating rate}
\label{subsec:heating and thermalization power law}
The previous solution can be trivially generalized to the case in which both the specific heating rate and the thermalization efficiency follow a power-law evolution:
\begin{equation}
\dot{\epsilon} = \dot{\epsilon}_0 \left( \frac{t}{t_0} \right)^{-\alpha} \, , \quad f_{\rm th} = f_{\rm th,0} \left( \frac{t}{t_0} \right)^{-\beta} \, .
\end{equation} 
In this case, \refeq{eq:phi_n sol 1}-\refeq{eq:A_n sol 1} are still a solution of \refeq{eq:temporal_eq}, once $\alpha$ has been replaced by $\alpha'$ and $\alpha' \equiv \alpha + \beta$.

\subsubsection{Constant opacity, constant thermalization efficiency and power-law heating rate with exponential terms}
\label{subsec:heating power law and exp}

We then consider the case in which the opacity and the thermalization efficiency are constant in time, while the specific heating rate can be written as
\begin{equation}
\dot{\epsilon} = \dot{\epsilon}_0 \left( \frac{t}{t_0} \right)^{-\alpha} + B e^{-t/\beta} \, .
\label{eq: heating w exponential}
\end{equation}
The solutions of \refeq{eq:temporal_eq} becomes:
\begin{equation}
  \begin{split}
     \phi_n(t) = & \exp\left(-\frac{\pi^2n^2t^2}{2t_0\tau_0} \right)\left[K_n+A_n\Gamma\left(1-\frac{\alpha}{2},-\frac{\pi^2n^2t^2}{2t_0\tau_0}\right) + \right. \\    
     & + \left. B_n {\rm erfi} 
     \left( \frac{t_0 \tau_0 - \pi^2 n^2 \beta t}{\sqrt{2} \pi n \sqrt{t_0 \tau_0} \beta} \right)  \right]+ C_n e^{-t/\beta} \, ,
  \end{split}
\end{equation}
where ${\rm erfi}(x) \equiv -i~{\rm erf}(ix)$ is the imaginary error function and the error function is defined as
\begin{equation}
{\rm erf}(z) \equiv \frac{2}{\sqrt{\pi}} \int_0^z e^{-t^2} {\rm d}t \, .
\end{equation}
The $B_n$, $C_n$ and $K_n$ coefficients read
\begin{equation}
B_n \, = \, \left(-1 \right)^n \exp\left(-\frac{t_0 \tau_0}{2 \pi^2 n^2 \beta^2} \right) \frac{\sqrt{t_0} \tau_0^{3/2} \rho_0 B}{\pi^{7/2}E_0 \beta n^4} \, ,\\
\end{equation}
\begin{equation}
C_n \, = \, \left(-1 \right)^{n+1} \frac{\sqrt{2} \tau_0 \rho_0 B}{\pi^3 E_0 n^3} \, ,\\
\end{equation}
\begin{equation}
    \begin{split}
    K_n \, = \, \, \, & \delta_{n,1} \exp\left(-\frac{\pi^2 n^2 t_0}{2 \tau_0} \right) - 
    A_n \Gamma\left(1-\frac{\alpha}{2},-\frac{\pi^2n^2t_0}{2 \tau_0}\right) + \\ 
     & - B_n {\rm erfi} \left( \frac{t_0 \tau_0 - \pi^2 n^2 \beta t_0}{\sqrt{2} \pi n \sqrt{t_0 \tau_0} \beta} \right) - C_n \exp\left( \frac{\pi^2 n^2 t_0}{2 \tau_0}-\frac{t_0}{\beta} \right) \, .
    \end{split}
\end{equation}    
This solution can be trivially generalized to the case in which more than one exponential term is added to the power law term in \refeq{eq: heating w exponential}.

\subsubsection{Power-law opacity, thermalization efficiency and heating rate}
\label{subsec:heating, thermalization and opacity power law}
Finally, we consider the case in which the opacity, the thermalization efficiency and the specific heating rate have a temporal power-law dependence: 
\begin{equation}
\dot{\epsilon} = \dot{\epsilon}_0 \left( \frac{t}{t_0} \right)^{-\alpha} \, , \quad f_{\rm th} = f_{\rm th,0} \left( \frac{t}{t_0} \right)^{-\beta} \, , \quad \kappa = \kappa_0 \left( \frac{t}{t_0} \right)^{-\gamma}  \, .
\end{equation} 
In this case, $\tau$ can be expressed as $\tau = \tau_0 (t/t_0)^{-\gamma}$
so that \refeq{eq:temporal_eq} becomes:
\begin{equation}\label{eq:temporal_eq_opacity_dep}
   \phi_n'(t)+\left(\frac{t}{t_0}\right)^{1+\gamma}
   \frac{(n^2\pi^2)}{\tau_0}\phi_n(t) =(-1)^{n+1} \frac{\rho_0 \sqrt{2} \dot{\epsilon}_0 f_{\rm th,0}}{n\pi E_0 } 
     \left(\frac{t}{t_0}\right)^{1-\alpha'}  
    \, .
\end{equation}
In this case, the solution of \refeq{eq:temporal_eq_opacity_dep} becomes:
\begin{equation}
    \begin{split}
    \phi_n(t) = & \exp\left({-\frac{\pi^2n^2t^{2+\gamma}}{(2+\gamma)t_0^{(1+\gamma)}\tau_0}} \right) \times \\ 
    & \times \left[K_n+A_n\Gamma\left(\frac{2-\alpha'}{2+\gamma},-\frac{\pi^2n^2t^{2+\gamma}}{(2+\gamma)t_0^{1+\gamma}\tau_0}\right)\right] \, ,
    \end{split}
    \label{eq:phi_n sol 3}
\end{equation}
where $A_n$ and $K_n$ are defined as
\begin{equation}
    A_n= (-1)^{n} \frac{\sqrt{2} \rho_0 f_{\rm th,0} \dot{\epsilon}_0 t_0}{E_0 n \pi (\gamma + 2)} 
    \left( - \frac{t_0}{\tau_0} \frac{ n^2 \pi^2}{(\gamma+2)} \right)^{\frac{\alpha'-2}{\gamma+2}}
    \label{eq:A_n sol 3} 
\end{equation}
and
\begin{equation}
K_n =  \delta_{n,1} \exp\left({\frac{\pi^2 n^2 t_0}{(2 + \gamma) \tau_0} }\right) - 
A_n \Gamma\left(\frac{2-\alpha'}{\gamma+2},-\frac{\pi^2n^2t_0}{(2+\gamma)\tau_0}\right) \, .
\end{equation}

\subsection{Optically thin ejecta treatment} \label{subsec:op_thin}
At the times relevant for the kilonova, we expect the r-process material outside of the ejecta photosphere to provide a non-negligible contribution to the heating powering the emission, especially when a proper photosphere will eventually not be identifiable anymore. However this contribution is expected to be considerably different with respect to the one provided by the same portion of the ejecta if the latter were optically thick, i.e. if we assumed $R_{\rm ph}(t)=R_{\rm max}(t)$.\\
Therefore, in addition to the radiation emitted at the photosphere, we approximate the bolometric luminosity $L_{\rm thin}(t)$ from the thin region outside of it, following the prescription of \citet{Grossman:2013lqa,Martin:2015hxa,Wu:2018mvg}. We thus divide this region into $N_{\rm thin}$ layers of equal mass d$M_i$ assuming local thermodynamics equilibrium within each layer, and we express such contribution as:
\begin{equation}
    L_{\rm thin}(t)=\sum_{i=1}^{N_{\rm thin}} f_{{\rm th},i}(t)\dot{\epsilon}(t){\rm d}M_i \, ,
\end{equation}
where the sum runs over the discrete thin shells, while $\dot{\epsilon}(t)$ and $f_{\rm th,i}(t)$ are the specific radioactive heating rate and the space-dependent binned thermalization efficiency, respectively, as described in \refsec{subsec:heating_rates}. 

\begin{figure}
\includegraphics[width=\columnwidth]{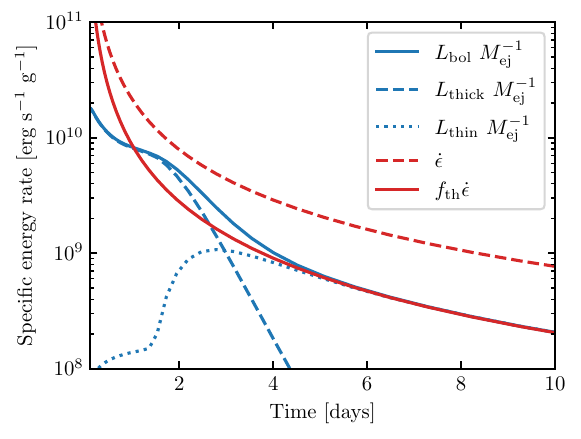}
\caption{Bolometric luminosity per unit of ejecta mass and its two contributions from the optically thick and thin ejecta as a function of time, for a spherically symmetric model with $M_{\rm ej}=0.01$ $M_{\odot}$, $v_{\rm ej}=0.2~c$ and $\kappa=5$ ${\rm cm^2~g^{-1}}$. Also shown in red is the specific radioactive heating rate powering the emission and its thermalized fraction. The heating curve was obtained following the procedure described in \refsec{subsec:heating_rates} using an entropy of $s=10$ $k_{\rm B}~{\rm baryon^{-1}}$ and an expansion timescale of $\tau_{\rm exp}=5$ ms.}
\label{fig:luminosity}
\end{figure}
The total bolometric luminosity of the ejecta is simply obtained by summing the contributions from the two separate regions:
\begin{equation}
    L(t)=L_{\rm thick}(t)+L_{\rm thin}(t) \, .
\end{equation}
In \reffig{fig:luminosity}, the total luminosity is displayed together with its components for a simple spherically symmetric model. As visible, the early light curve is dominated by the thick ejecta, which constitutes the majority of the total mass. In this phase most of the energy provided by the radioactive decays is trapped within the ejecta due to the high optical depths. After a few days, the ejecta density has decreased enough for this energy to escape, enhancing the emission up to be instantaneously greater than the thermalized heating rate. Meanwhile, a second contribution to the luminosity steps in, as a relevant portion of optically thin mass emits radiation as well. Finally, after several days, the thick bulk of the ejecta disappears, and the luminosity is completely determined by the optically thin matter. Since the latter is transparent to thermal radiation, the thermalized decay energy escapes without further processing, and the luminosity is equal to the thermalized heating rate. However, the latter is now only a small fraction of the total decay energy rate, since the lower densities make the thermalization process inefficient. Thus, the thermal emission will eventually fade away.\\

Along with the temperature of the photosphere $T_{\rm ph}$, we want to characterize also the temperature in the layers outside of it. For this purpose, we assign to each bin a temperature $T_i(t)$ on the basis of the radial profile proposed by \citet{Wollaeger:2017ahm} and derived for a radiation dominated ideal gas using \refeq{eq:density_profile}:
\begin{equation}
    T(t,x)=T_0(t)\left(1-x^2\right) \, .
\end{equation}
For each time, we fix the factor $T_0(t)$ by requiring the continuity of the profile with the photospheric temperature $T_{\rm ph}(t)$. Therefore, for every bin $i$ we obtain:
\begin{equation}
    T_i(t)=T_{\rm ph}(t)\frac{1-x_i^2}{1-x_{\rm ph}^2(t)} \, ,
\end{equation}
where $x_i$ is the position of the bin and $x_{\rm ph}(t)=R_{\rm ph}(t)/R_{\rm max}(t)$ is the position of the photosphere.\\
Recently, \citet{Pognan:2021wpy} showed that a synthetic non-local thermodynamics equilibrium evolution of the temperature in the late expanding ejecta features a re-increase from a minimum reached around a few tens of days post merger. This result was obtained by taking into account the material excitation and ionization states in a more careful calculation of the ejecta heating and cooling processes, using the spectral synthesis code \texttt{SUMO}.
In light of the above computations, we expect the temperature to remain roughly constant at the late times still relevant for the first kilonova phase. Therefore, in order to describe the thermal emission, here we find sufficient to set a unique time-independent minimum temperature value $T_{\rm floor}$ for both the thick and the thin part of the ejecta.

\section{Multi-component anisotropic semi-analytic kilonova model} \label{sec:xkn framework}

Ejecta from compact binary mergers are expected to occur in different components, characterized by different properties. Moreover, the ejection mechanisms can result in a anisotropic structure of the ejecta. Motivated by this, we set up a multicomponent, anisotropic kilonova framework.
In particular, we closely follow the set-up first proposed by \citet{Perego:2017wtu}, originally based on \citet{Martin:2015hxa} and reprised by \citet{Barbieri:2019sjc,Barbieri:2019kli,Camilletti:2022jms}.

The framework assumes axial symmetry around the rotation axis of the binary (denoted as $z$), as well as reflection symmetry about the $z=0$ plane. The polar angle $\theta$ is discretised in a series of $N_{\theta}$ bins which can be equally spaced either in the angle itself or in $\cos{\theta}$. A kilonova model is specified once the polar distribution of all the relevant quantities (i.e. mass, velocity, opacity or electron fraction, entropy, expansion time scale) are given for each of the ejecta components. Inside each angular bin and for each component, the radial kilonova model described in \refsec{sec:1d_model} (or alternatively the model from \citet{Grossman:2013lqa}) can be employed to compute the contribution to the luminosity emerging from that angular bin. 
Being an extensive quantity, the mass inside the bin needs to be scaled by the factor $4\pi / \Delta \Omega$, where $\Delta \Omega$ is the bin solid angle. All the other input quantities are otherwise intensive and do not need any rescaling.
Once computed, the isotropic luminosity resulting from the 1D model is scaled again based on the actual emitting solid angle, i.e. it is multiplied by $\Delta \Omega/4\pi$. 
Within the same angular bin and in the presence of more than one ejecta component, the corresponding luminosity contributions are summed together, assuming that photons emitted from the innermost components irradiate the outermost ones and are subsequently re-processed and re-emitted on a time scale smaller than the expansion one. Moreover, at each time we locate the photosphere of the overall ejecta at the position of the larger individual photosphere.
This approach assumes that the different components are nested and that they do not cross each other significantly during the kilonova emission. We expect these hypotheses to be appoximately verified once the homologous expansion phase has been reached and if the late time ejecta are systematically slower than the first expelled ones.

The present implementation includes characteristic analytic functional forms for the angular dependences: uniform distributions, step functions, $\sin{\theta}$, $\sin^2{\theta}$, $\cos{\theta}$ and $\cos^2{\theta}$ dependences. 
Despite their simplicity, some of these distributions were demonstrated to be remarkably valid in broadly reproducing the outcomes of general-relativistic hydrodynamical simulations, accounting for the preferential equatorial direction of the dynamical component, as well as the excursion in the electron fraction caused by high-latitudes neutrino irradiation.
Additionally, the code can interpolate on its angular grid arbitrary distributions, such as azimuthally-averaged angular profiles extracted from numerical simulations \citep[see, for example,][]{Camilletti:2022jms}.

Typical kilonova models employed in the past used up to three different components \citep[][see, e.g.,]{Perego:2017wtu,Breschi:2021tbm}. In the case of two components, the fastest one usually refers to the dynamical ejecta, while the slowest one to the disc wind ejecta of viscous origin. A third, intermediate component is sometimes used, possibly originated by magnetic- \citep[see e.g.][]{} or neutrino-driven wind components \citep[e.g.][]{Perego:2014qda}, as well as from spiral wave wind ejecta \citep[e.g.][]{Nedora:2019jhl,Nedora:2020pak}.

In the source frame, the emission is assumed to be thermal and the spectral fluxes are described by a Planckian spectral distribution $B_{\nu}(T)$, i.e.:
\begin{equation}
    B_{\nu}(T)=\frac{2\pi h\nu^3}{c^2}\frac{1}{{\rm exp}\left(\frac{h\nu}{k_{\rm B}T}\right)-1} \, ,
\end{equation}
with $k_{\rm B}$ the Boltzmann constant and $h$ the Planck constant, both at the photosphere as well as within each thin external layer. In the former case, $T$ is the photospheric temperature, while in the latter it is the temperature inside each mass shell. 

If the source is located at a luminosity distance $D_{L}$, corresponding to a redshift $z$, for an observer on Earth characterized by a viewing angle $\theta_{\rm view}$, the radiant flux at frequency $\nu$ and time $t$ (measured in the observer frame) will be the sum over the angular bins of the contributions from the thick ejecta $F^{\rm thick}_{\nu,k}$ and the thin ejecta $F^{\rm thin}_{\nu,k}$ (computed in the source frame), once the redshift correction has been applied to the time, frequency and luminosity:
\begin{eqnarray}\label{eq:obs_flux}
    f_{\nu}(t) & = & \frac{(1+z)}{4\pi  D_L^2}\sum_{k=1}^{N_\theta} \left\{ p_k(\theta_{\rm view})\frac{4\pi}{\Delta \Omega_k}\left[ F^{\rm thick}_{(1+z)\nu,k} \left( \frac{t}{1+z} \right) \right. \right. \nonumber \\
    & & \left. \left.+ F^{\rm thin}_{(1+z)\nu,k}\left( \frac{t}{1+z} \right)\right] \right\} \, ,
\end{eqnarray}
with:
\begin{equation}
    F^{\rm thick}_{\nu,k}(t')=\frac{L_{{\rm thick},k}(t')}{\sigma_{\rm SB} T_{{\rm ph},k}^4(t')}B_{\nu}(T_{{\rm ph},k}(t')) \, ,
\end{equation}
and:
\begin{equation}
    F^{\rm thin}_{\nu,k}(t')=\sum_i\frac{f_{{\rm th},i,k}(t')\dot{\epsilon}_{k}(t'){\rm d}M_{i,k}}{\sigma_{\rm SB} T_{i,k}^4(t')}B_{\nu}(T_{i,k}(t')) \, .
\end{equation}
where $L_{\rm thick,k}$ is the photospheric luminosity of the bin $k$, characterized by a photospheric temperature $T_{{\rm ph},k}$. 
The factors $p_k(\theta_{\rm view})$ in \refeq{eq:obs_flux} account for the effective emission area as seen by the observer \citep{Martin:2015hxa}, and are calculated using the formula:
\begin{equation}
    p_k(\theta_{\rm view})=\frac{1}{\pi}\int_{\Vec{q}(\theta_{\rm view})\cdot\Vec{n}_k>0}\Vec{q}(\theta_{\rm view})\cdot{\rm d}\Vec{\Omega} \, ,
\end{equation}
where $\Vec{q}(\theta_{\rm view})$ is the unit vector in the observer direction, while $\Vec{n}_k$ is the unit vector pointing radially outwards from the surface of the bin $k$.\\
Finally, we compute the AB magnitude at a photon frequency $\nu$ as:
\begin{equation}
    m_{\rm AB,\nu}(t)=-2.5~{\rm log}_{10}(f_{\nu}(t))-48.6 \, .
\end{equation}

\section{Input physics} \label{sec:input_physics}

\subsection{Heating rates}\label{subsec:heating_rates}
The heating rate powering the kilonova originates from the many decays of heavy elements produced in the r-process nucleosynthesis, and as such it can be computed by employing a nuclear reaction network. The latter calculates the time evolution of the nuclides abundances while keeping track of the energy released in the process. Results obtained by nuclear network calculations retain a strong dependence on the properties of the ejecta, and in particular on the entropy, electron fraction and expansion timescale at the freeze-out from nuclear statistical equilibrium (NSE) \citep[see, e.g.,][]{Hoffman:1996aj,Lippuner:2015gwa}. Furthermore, nuclear network calculations also depend on the nuclear physics employed, e.g. on the choice of the theoretical nuclear mass model, the reaction rates or the fission fragment distribution. This sensitivity is particularly strong at low electron fractions and the nuclear physics uncertainties can lead to changes in the predicted heating rates of about one order of magnitude \citep{Mendoza-Temis:2014,Rosswog:2016dhy,Zhu:2020eyk}.\\

\begin{figure*}
\includegraphics[width=\textwidth]{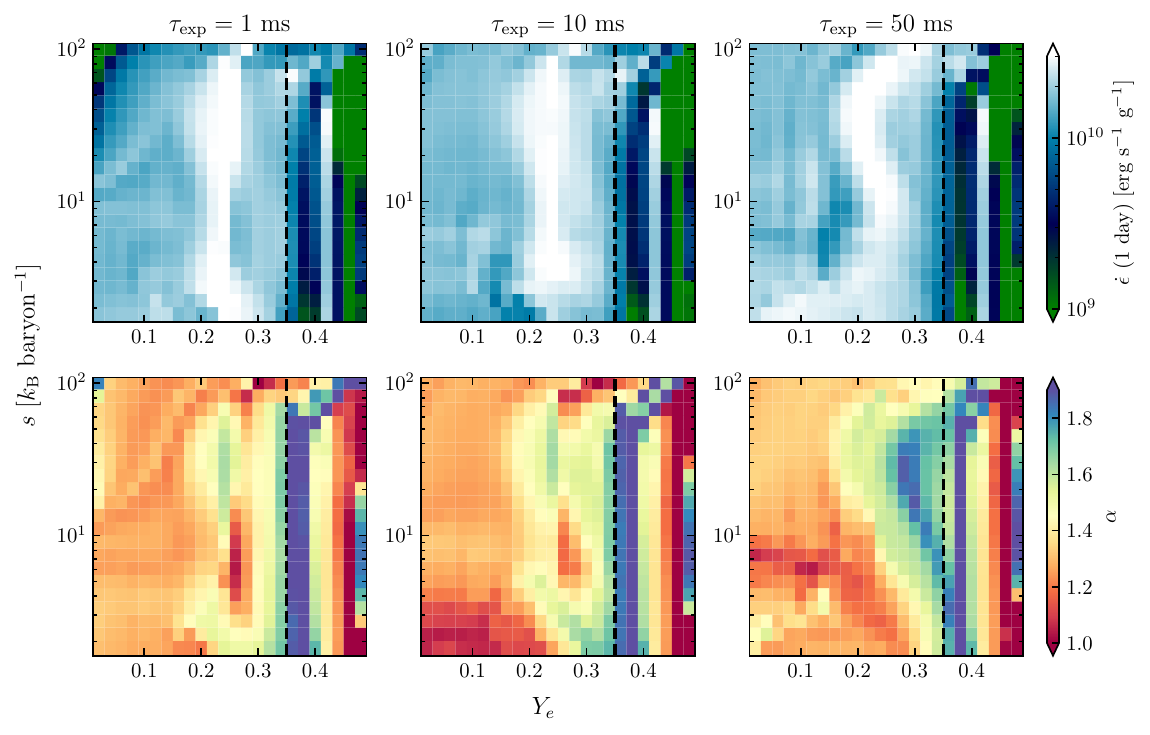}
\caption{Heating rate fit parameters as functions of the initial electron fraction $Y_e$ and entropy $s$, for fixed values of the expansion timescale $\tau_{\rm exp}$.
The dashed line separates the region where we interpolate the parameters linearly (left) from the region where the continuity of the parameters is poor and we use a nearest-point interpolation (right).}\label{fig:epsfit_par}
\end{figure*}
With the purpose to provide our kilonova model with a heating rate valid for arbitrary initial conditions, we consider the results of the broad nucleosynthesis calculations reported in \citet{Perego:2020evn}. In that work, the nuclear composition evolution of a set of Lagrangian fluid elements is computed using the nuclear reaction network \texttt{SkyNet} \citep{Lippuner:2017tyn} with the finite-range droplet macroscopic nuclear mass model (FRDM) \citep{Moller:2015fba}. Each \texttt{SkyNet} run is initialized at a temperature of $6$ GK in NSE, and identified by the values of the initial electron fraction $Y_e$, entropy $s$, and expansion timescale $\tau_{\rm exp}$. The latter are considered as initial parameters and later evolved consistently by the network. More details about these nucleosynthesis calculations can be found in \citet{Perego:2020evn}.
In particular, the heating rates we employ are computed over a comprehensive grid of $\sim11700$ distinct trajectories with $0.01\leq Y_e\leq0.48$ linearly spaced, $1.5$ $k_{\rm B}~{\rm baryon^{-1}}$ $\leq s\leq 200$ $k_{\rm B}~{\rm baryon^{-1}}$ and $0.5$ ms $\leq\tau_{\rm exp}\leq200$ ms logarithmically spaced. These intervals are expected to bracket the properties of the ejecta from BNS and NSBH mergers.
We fit the heating rate trajectories obtained with \texttt{SkyNet} over the time interval $0.1$ days $\leq t\leq50$ days after merger, using the following power-law dependence:
\begin{equation}\label{eq:heating_rate}
    \dot{\epsilon}=\dot{\epsilon}_{\rm 1d} \left( \frac{t}{\rm 1~day} \right)^{-\alpha} \, ,
\end{equation}
where $\dot{\epsilon}_{\rm 1d}$ and $\alpha$ are fit parameters, with typical values $\alpha \sim1.3$ and $\dot{\epsilon}_{\rm 1d}\sim10^{10}$ ${\rm erg~s^{-1}~g^{-1}}$. Such a temporal dependence in the heating rate is expected from the decay of large sample of unstable nuclei \citep{Metzger:2010sy,Korobkin:2012uy}. Moreover, it is equivalent to \refeq{eq:heating rate KN model}, one of the functional forms used in the optically thick kilonova model described in \refsec{sec:1d_model}, provided a conversion factor between the fit reference time (1 day) and $t_0$.
The quality of each single fit is evaluated using a mean fractional log error as employed in \cite{Lippuner:2015gwa}, defined as:
\begin{equation}
\Delta(\dot{\epsilon})=\left<\frac{|\ln(\dot{\epsilon}^o(t))-\ln(\dot{\epsilon}(t))|}{\ln(\dot{\epsilon}^o(t))}\right> \, ,
\end{equation}
where $\dot{\epsilon}^o(t)$ is the original \texttt{SkyNet} heating rate trajectory, while the mean is performed over the fit time window without weighing over the time steps, in order to account for the original \texttt{SkyNet} resolution. For most trajectories we find the average relative errors to be smaller than $\sim1\%$. The largest errors are found at the boundary of the \texttt{SkyNet} grid, where the relative error can be as large as $\sim 5\%$.\\
In \reffig{fig:epsfit_par}, the values of the fitting coefficients are plotted against $Y_e$, $s$ and $\tau_{\rm exp}$ for representative sections of the \texttt{SkyNet} grid. As shown in the left column, for a fixed $Y_e$ the fit parameters are generally smooth functions of the two other thermodynamic variables, and in particular the value of $\alpha$ remains roughly constant (for $Y_e\lesssim0.2$ it hardly deviates from $\sim1.3$, as already found in \citet{Korobkin:2012uy}), while the value of $\dot{\epsilon}_0$ varies within a factor of a few. On the other hand, the variability of the fit parameters increases as the electron fraction is left free to vary. This strong and non-trivial dependence of the heating rate on the electron fraction is more evident for high $Y_e$ values, where the radioactive heating can be dominated by the decay of individual nuclear species, depending on the specific ejecta conditions. However, we find that the continuity in the fit parameters endures at least in the region which is more relevant to our study, i.e. for $Y_e\leq0.36$, $s\leq90$ $k_{\rm B}~{\rm baryon^{-1}}$ and $\tau_{\rm exp}\leq30$ ms. We therefore adopt a trilinear interpolation of the fitting coefficients as functions of $Y_e$, $s$, and $\tau_{\rm exp}$ in that region, while isolated points or boundary areas for which the continuity of the fitting coefficients is poor are treated by using a nearest-point interpolation. 

\begin{figure}
\includegraphics[width=\columnwidth]{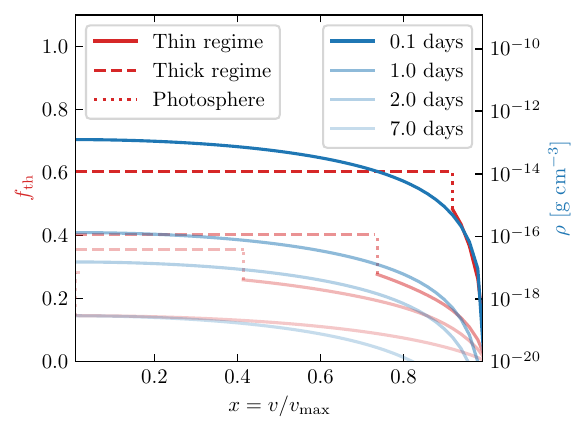}
\caption{Constructed thermalization efficiency radial profile at different days post-merger. The ejecta density radial profile from \refeq{eq:density_profile} is shown as reference.}
\label{fig:thermalization}
\end{figure}
In order to account for the efficiency with which decay products thermalize in the ejecta, we apply a thermalization efficiency factor to the heating rate as follows.
For the thick core of the ejecta, we consider both a constant thermalization efficiency (compatible with all the analytic solutions presented in \refsec{sec:1d_model}) and a thermalization efficiency with a time evolution $f_{\mathrm{th}} = f_{\rm th,0}\left( t/t_0 \right)^{-\beta}$, as described in \refsec{subsec:heating and thermalization power law}, and \refsec{subsec:heating, thermalization and opacity power law}.
The latter formula approximately mimics the decreasing in the thermalization behaviour expected during the first day in the optically thick ejecta. The values of $f_{\rm th,0}$ and $\alpha_{\rm th}$ can be fixed by imposing, for example, a thermalization efficiency of $\sim 0.7$ and $0.4$ at $0.1$ days and $1$ day, respectively. For the thin layers of the ejecta instead, we model a thermalization efficiency profile starting from the analytic formula proposed in \citet{Barnes:2016umi} and fitted on the properties of the ejecta:
\begin{equation}\label{eq:thermalization}
    f_{\mathrm{th}}(t,x)=0.36\left[\exp(-aX)+\frac{\ln(1+2bX^d)}{2bX^d}\right] \, , 
\end{equation}
where $a$, $b$ and $d$ are the fit parameters. In that work, this expression was obtained by assuming \refeq{eq:density}, and $X(t,x)=t$. Here instead, we adopt \refeq{eq:density_profile}, and $X(t,x)=t\left(1-x^2\right)^{-1}$. We interpolate the fit parameters in \refeq{eq:thermalization} on the tabulated grid reported in \citet{Barnes:2016umi}, which spans the intervals $1\times10^{-3}~M_{\odot} < M_{\mathrm{ej}} < 5\times10^{-2}~M_{\odot}$ for the total ejecta mass and $0.1~c < v_{\mathrm{ej}} < 0.3~c$ for its average velocity.
This combination of different efficiencies is motivated by the fact that, on one hand, we expect the decay energy in the thick bulk to thermalize in a similar way as long as the density is sufficiently high. In particular, roughly $\sim35\%$ of the energy escapes in the form of neutrinos, $\sim45\%$ is constituted by $\gamma$-rays which efficiently heat the material only within the first day post-merger, and the remaining $\sim20\%$ is carried by $\beta$-particles, $\alpha$-particles and fission yields \citep{Barnes:2016umi}. On the other hand, the thermalization efficiency drops significantly in the outer layers of the ejecta, where the lower density makes it harder for the decay products to deposit their energy through thermal processes. 
\reffig{fig:thermalization} shows the modeled thermalization efficiency profile for different times after merger. By construction, the efficiency in the thin ejecta rapidly declines to values $<20\%$ after a few days post merger. Concurrently, the photosphere radius receeds inward until it disappears. Despite being an artifact, the discontinuity in the efficiency profile at the photosphere radius is not inconsistent with our photosphere model, which assumes a sharp difference in the properties of matter between the thick and the thin ejecta regions. However, we acknowledge the crudeness of the overall thermalization treatment, which does not rigorously account for the dependency on the ejecta conditions of the specific deposition processes involved. Therefore, we leave for a future investigation the impact of more detailed thermalization descriptions on the resulting kilonova light curves.

\subsection{Opacities}\label{subsec:opacities}
\begin{figure}
\includegraphics[width=\columnwidth]{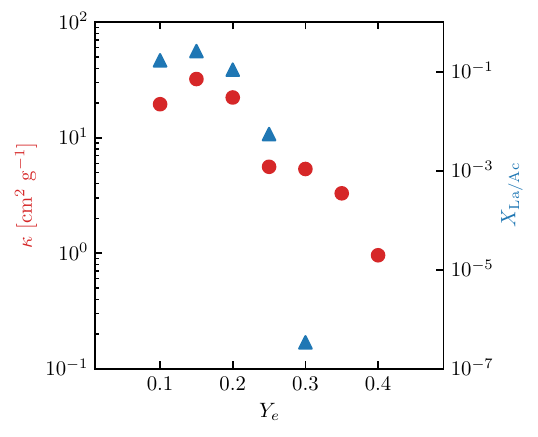}
\caption{Grey opacity values derived by \citet{Tanaka:2019iqp} for temperatures $5000$ K $<T<10000$ K and densities $\rho\sim10^{-13}~{\rm g~cm^{-3}}$ and for different ejecta compositions characterized by a specific value of electron fraction. The mass fraction of Lanthanides and Actinides is reported for every considered composition.}
\label{fig:opacity}
\end{figure}
In our framework, we can consider the opacity for the r-process material in the ejecta as a free parameter of the model. Alternatively, we can also provide composition-dependent opacity values following the work of \citet{Tanaka:2019iqp}, in which systematic atomic structure calculations on each element between Fe ($Z=26$) and Ra ($Z=88$) are performed using the integrated code \texttt{HULLAC} \citep{2001JQSRT..71..169B}. That study mainly focuses on the ejecta conditions around $1$ day after the merger, where the temperature is low enough ($T\lesssim20000$ K) to find the heavy elements ionization stages typically between I-IV. At this time, the density is assumed to be $\rho=1\times10^{-13}~{\rm g~cm^{-3}}$ (which is a typical value for an ejecta with mass $M_{\rm ej}\sim0.01~M_{\odot}$ and velocity $v_{\rm ej}\sim0.1~c$), and from here on the opacity in the IR, optical and UV is dominated by bound-bound transitions \citep{Kasen:2013xka}.\\
Bound-bound opacities on a fixed wavelength grid for the homologously expanding material are computed using the widely employed expansion opacity formalism:
\begin{equation}
    \kappa(\lambda)=\frac{1}{ct\rho}\sum_l\frac{\lambda_l}{\Delta\lambda}(1-e^{-\tau_l}) \, ,
\end{equation}
where the sum runs over all the transitions in the RT simulations within the bin $\Delta\lambda$. Planck mean opacities are then computed for a representative ejecta model with different mixtures of heavy elements, characterized by the value of the initial electron fraction $Y_e$ (see \refsec{subsec:heating_rates}).\\
In \reffig{fig:opacity} we report the grey opacity values derived by \citet{Tanaka:2019iqp} for ejecta temperatures of $5000$ K $<T<10000$ K, whereas a stronger temperature dependence is found for $T<5000$ K.
We note that in more recent works \citep[see, e.g.,][]{Banerjee:2023gye,Banerjee:2022doa,Banerjee:2020myd} such opacity calculations are extended to the ionization stages V-XI of the elements up to Ra, which are expected to be present for ejecta temperatures up to $\sim10^5$ K at times shorter than 1 day post-merger.
We therefore leave the corresponding suggested grey opacities as a possible alternative to the \citet{Tanaka:2019iqp} dataset.\\
In general, around 1 to a few days, if the electron fraction is low enough ($Y_e\lesssim0.25$), the grey opacity is dominated by lanthanides and actinides, with values $\kappa\gtrsim10~{\rm cm^2~g^{-1}}$. Instead, an increase in the electron fraction between $0.25\lesssim Y_e\lesssim0.35$ causes a general decrease of the opacity to values $\kappa\sim1-10~{\rm cm^2~g^{-1}}$, as the fraction of $f$- valence shell elements present in the ejecta decreases, leaving room for the $d$-shell atoms to provide the leading contribution. Finally, at even higher electron fractions $Y_e\gtrsim0.4$, the contribution from Fe-like elements dominates the opacity, which reaches values $\kappa\sim0.1-1~{\rm cm^2~g^{-1}}$.
In this instance, we interpolate the values in \reffig{fig:opacity} to uniquely determine the ejecta opacity on the basis of the input $Y_e$.

\section{Comparison with radiative transfer calculations}\label{sec:RT_comparison}

\subsection{Radiative transfer code}
In order to assess the level of reliability of our model, we set up a comparison between the light curves obtained by our semi-analytical model and the ones obtained by a RT kilonova simulation.
For the latter, we refer to \citet{Kawaguchi:2018ptg,Kawaguchi:2020vbf}, who employ the wavelength-dependent Monte Carlo RT code originally presented in \citet{Tanaka:2013ana}.
For a given density structure and abundance distribution, the code computes the time evolution of the photon spectrum in the UVOIR wavelength range, together with multicolor light curves. Differently from the first 3D version, \citet{Kawaguchi:2018ptg,Kawaguchi:2020vbf} assume the ejecta to be axisymmetric.
This allows for an increase in the simulation spatial grid resolution, and for the inclusion of special-relativistic effects in the photon transport.
Photon-matter interaction is described by considering elastic scattering off electrons, together with free-free, bound-free and bound-bound transitions. The contribution to the opacity from the latter is computed using the expansion opacity formalism described in \refsec{subsec:opacities}, while the atomic transition line list employed in the code is the one already used in \citet{Tanaka:2017qxj,Tanaka:2019iqp}. Since these atomic data concern the ionization stages I-III, the code is used only for temperatures up to $\sim10000$ K, below which further ionization stages are subdominant. Nuclear heating rates and elemental abundances are directly imported from the nucleosynthesis calculations of \citet{Wanajo:2014wha}, based on the post-processing of Lagrangian tracer particles obtained by a fully general relativistic simulation of a BNS merger with approximate neutrino transport. Each reaction network calculation starts from a representative thermodynamic trajectory with an initial electron fraction in the range $Y_e=0.09-0.44$. The fraction of thermalized energy is computed using the analytic formulae reported in \citet{Barnes:2016umi} for the different decay products. These formulae depend on the mass and velocity of the ejecta in a similar fashion to \refeq{eq:thermalization}. In particular, while the velocity parameter in the thermalization formulae is fixed to $v_{\rm ej}=0.3~c$, the mass parameter is set starting from the local density and considering a uniform sphere of radius $v_{\rm ej}t$.\\

\begin{figure}
\includegraphics[width=\columnwidth]{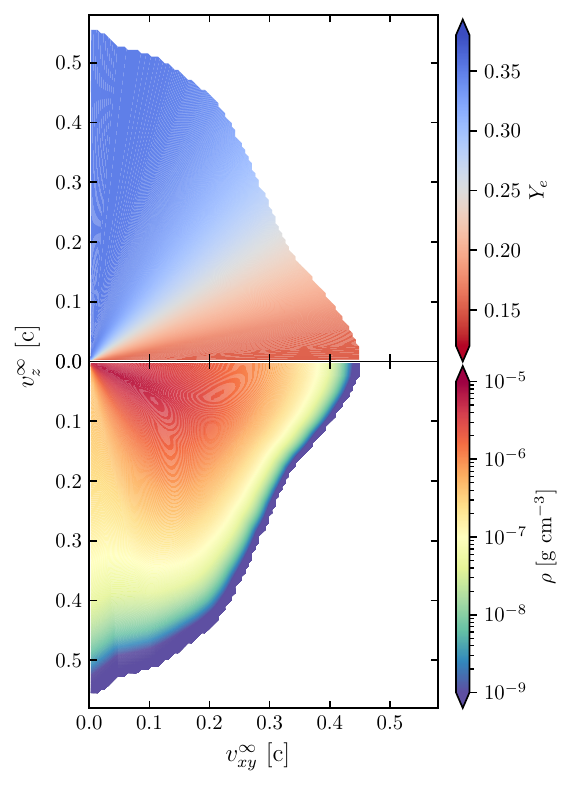}
\caption{Density and electron fraction distributions in the velocity space for the dynamical ejecta at $t=100$ s post-merger. The density radial profile follows the analytic prescription of \refeq{eq:density_profile}, while a radially constant electron fraction is assumed. Angular distributions are extracted from the GRHD simulation of an equal-mass BNS system with masses $M_{1,2}=1.364~M_{\odot}$ using the DD2 EOS \citep{Perego:2019adq}.}
\label{fig:NR_profiles}
\end{figure}

\begin{table*}
\caption{Fit parameters boundaries considered in the fitting procedure. The wider intervals adopted for the parameters associated to the secular ejecta reflect the major variability in composition of such component, based on the outcome of different numerical simulations.}
\centering
\begin{tabular}{l | c c | c c c c}
\hline
& \multicolumn{2}{c}{Secular wind} & \multicolumn{4}{c}{NR dynamical ejecta} \\
\cline{2-7}
& $\kappa$ [${\rm cm^2~g^{-1}}$] & $T_\mathrm{floor}$ [K] & $\kappa_\mathrm{high~lat}$ [${\rm cm^2~g^{-1}}$] & $\kappa_\mathrm{low~lat}$ [${\rm cm^2~g^{-1}}$] & $T_\mathrm{Ni}$ [K] & $T_\mathrm{LA}$ [K] \\
\hline
min & 0.5 & 0 & 0.1 & 5 & 0 & 0 \\
max & 50 & 6000 & 10 & 50 & 6000 & 3000 \\
\hline
\end{tabular}
\label{tab:priors}
\end{table*}

\subsection{Comparison setup}

We prepare our comparison by setting the same ejecta properties in both codes.
We consider two ejecta configurations, namely a lighter anisotropic dynamical component and a more massive spherically symmetric secular component.
This choice is motivated by the general necessity of modelling multiple components of matter ejection, which are required in order to reproduce the color bands of observed kilonovae, as in the case of AT2017gfo \citep{Cowperthwaite:2017dyu,Tanvir:2017pws,Tanaka:2017qxj}.
Regarding the secular component, we assume a total mass of $M_{\rm sec}=2.64\times10^{-2}~M_{\odot}$, an average velocity of $v_{\rm rms}=0.06~c$ and constant values for the electron fraction and specific entropy, i.e. $Y_e=0.2$ and $s=10$ ${\rm k_B~baryon^{-1}}$.
We compute the associated expansion time scale as $\tau_{\rm exp}=c/v_{\rm rms} \approx 17~{\rm ms}$.
These values are representative of the outcomes of simulations that investigate the evolution of disks emerging as remnants of compact binary mergers and accreting onto the central object.
In these simulations, a fraction between $\sim20-40\%$ of the disk mass is expelled during the secular evolution, with the initial disk mass $M_{\rm disk}\sim10^{-4}-10^{-1}~M_{\odot}$ \citep[see, e.g., ][]{Fahlman:2022jkh,Fujibayashi:2020dvr,Fernandez:2018kax,Hotokezaka:2012ze}.
Instead, for the dynamical component, we use the properties of the dynamical ejecta extracted from one GRHD simulation of a BNS merger with M0 neutrino transport approximation, chosen among those performed by \citet{Perego:2019adq} and compatible with the GW170817 event.
Despite the simulations considered in that work include different EOSs, they all lead to similar ejecta angular distributions, and therefore we arbitrarily select the simulation employing the HS(DD2) EOS \citep{Hempel:2009mc}.
The dynamical ejecta are identified with the matter unbound within the end of the simulation according to the geodesic criterion, i.e. the matter for which $|u_t|\geq c$, with $u_t$ the time-component of the four-velocity.
For an equal-mass binary with masses $M_{1,2}=1.364~M_{\odot}$, a total dynamical ejecta mass $M_{\rm dyn}=2.7\times10^{-3}~M_{\odot}$ was found.
The properties of this component are recorded as matter crosses a extraction spherical surface characterized by a coordinate radius $r_{\rm E}=294$ km, and are then reduced to an axisymmetric configuration by averaging over the azimuthal angle.
In particular, \texttt{xkn} is informed with the angular distributions of the ejecta mass, average electron fraction and entropy, and average velocity at infinity, calculated as $v_{\rm rms}^\infty=c\sqrt{1-(c/u_t)^2}$.
We choose the profile given by \refeq{eq:density_profile} to describe the radial density structure of each ejecta component both in the RT simulation and in \texttt{xkn}.
Moreover, we assume a radially constant electron fraction in order to fix the composition.
The resulting configuration of the dynamical ejecta as depicted in \reffig{fig:NR_profiles} reflects the general characteristics of this component as obtained in many merger simulations: neutron-rich matter is expelled preferentially across the equatorial plane partially through tidal forces, while shock-heated material subject to stronger neutrino irradiation and thus less neutron-rich escapes also at small polar angles.\\

The input profiles described above uniquely determine all the components of both the models, including energy deposition rates, elemental abundances and opacities, with the only exception of one remaining free parameter in \texttt{xkn}, that is the photospheric floor temperature $T_{\rm floor}$.
We remark that the employed radioactive heating rates, as well as the prescription for the thermalization efficiency, are not coincident between the two models, although derived from the same initial conditions.
However, the final energy deposition rates agree within a factor of a few, and we account for this discrepancy as being part of the general difference between the kilonova models.
Furthermore, we acknowledge that the opacity treatment in our \texttt{xkn} model is significantly approximated: in addition to the adoption of grey values, we assume the opacity to be constant in time or at most to evolve according to a power-law, when characterizing the ejecta through their entire evolution and depth.
In reality we expect the Planck mean opacity to vary by at least one order of magnitude between different regions and epochs.
Therefore, the adoption of the opacity values derived by \citet{Tanaka:2019iqp} and described in \refsec{subsec:opacities} is not more physically motivated than treating the opacity as a free parameter, and, for this reason, in the comparison we consider both possibilities.

The RT data employed in the comparison consist of the bolometric luminosity, $L^\mathrm{RT}_{\rm bol}(t)$, and of the AB magnitudes, $m^\mathrm{RT}_{\rm AB,\lambda,\theta}(t)$ at different wavelengths $\lambda$, observed from multiple viewing angles $\theta_{\rm view}\in[0^\circ,90^\circ]$.
We thus fit our free parameters to both sets of data separately, considering a logarithmically spaced time mesh, from $0.5$ to $15$ days.
Within this time frame, we assume that the assumptions of our model are better verified, and that the RT calculations are more reliable, whereas temperatures throughout the ejecta are well below $10000$ K, justifying the employed atomic data.\\
We define two error functions in order to establish the fit procedure. For the bolometric luminosity, we compute the absolute logarithmic error between our model luminosity, $L^\mathrm{M}_{\rm bol}$, and the RT result, $L^\mathrm{RT}_{\rm bol}$, averaged over all $N_t$ data points in the considered time frame:
\begin{equation}\label{eq:err_L}
    \mathrm{err}_L=\frac{1}{N_t}\sum_{t_i}\left|{\rm log}\left(\frac{L^{\rm M}_{\mathrm{bol},i}}{L^{\rm RT}_{\mathrm{bol},i}}\right)\right| \, .
\end{equation}
For the AB magnitudes, we consider three representative broadband filters, namely the $K$ ($\lambda=2157$ nm), $z$ ($\lambda=972$ nm) and $g$ ($\lambda=475$ nm) filters. The light curves are calculated assuming a source luminosity distance of $D_{L}=40$ Mpc, corresponding to the estimated distance for the merger associated to the AT2017gfo signal. Since our kilonova model is better suited to reproduce the light curve behaviour around the emission peak, only data points such that $m^{\rm RT}_{\mathrm{AB},i}<30$ are considered, in order to avoid having the fits influenced by too dim values. In a similar fashion to the bolometric luminosity, we compute the absolute error between the magnitudes across the three different wavebands and two different viewing angles, i.e. $0^\circ$ and $90^\circ$, averaging over all $6\times N_t$ data points in the same time interval:
\begin{equation}\label{eq:err_mag}
    \mathrm{err}_m=\frac{1}{6N_t}\sum_{t_i:m^{\rm RT}_{\mathrm{AB},i}<30} \left( \sum_{g,z,K} \left( \sum_{\theta=0^\circ,90^\circ}~|m^{\rm M}_{\mathrm{AB},i}-m^{\rm RT}_{\mathrm{AB},i}| \right) \right) \, .
\end{equation}
We perform two sets of runs, one for each ejecta configuration, i.e. one for the secular isotropic ejecta and one for the dynamical anisotropic ejecta.
Furthermore, for each set, we take into account two possibilities.
In one case we allow the opacity in our model to vary freely, and in particular for the anisotropic setup we assume it to follow a step function, i.e. we adopt a higher value, $\kappa_\mathrm{low~lat}$, at low latitudes ($\theta\gtrsim45^\circ$) in correspondence of a neutron-rich environment with $Y_e\lesssim0.25$, and a lower value, $\kappa_\mathrm{high~lat}$, at high latitudes ($\theta\lesssim45^\circ$) where $Y_e\gtrsim0.25$.
In the other case instead we compute the opacity using the $Y_e$ parametrization from \citet{Tanaka:2019iqp}, leaving us with only the photospheric floor temperature to be fitted. To be consistent with the opacity prescription, for the anisotropic ejecta setup we consider a $Y_e$-dependent floor temperature parameterized by the two values $T_{\rm Ni}$ and $T_{\rm LA}$.
In \reftab{tab:priors} we report the adopted ranges for the parameters included in the fit procedure.
Finally, each calculation is repeated using the semi-analytic kilonova model presented in \citet{Perego:2017wtu} for comparison purposes.
The latter shares the same multicomponent, anisotropic framework as the model presented in this work.
However, the underlying kilonova model is not based on the solution of the diffusion equation, but it is a phenomenological description based on timescale arguments, presented in
\citet{Grossman:2013lqa} and \citet{Martin:2015hxa}.
Due to this distinction, we name the previous model as \texttt{xkn-ts}, as opposed to our new \texttt{xkn-diff} model.

\subsection{Comparison results}

\begin{table*}
\caption{Best fit parameters obtained in the case of the bolometric luminosity (top) and magnitudes (bottom) with corresponding fit errors for the dynamical (right) and the secular (left) component configurations, as obtained by using \texttt{xkn-diff} and \texttt{xkn-ts} models, with and without a fixed opacity.
Cases for which the fit rails against the chosen boundaries are highlighted in gray.
(*) is used to indicate that the parameter value is fixed, either manually for the secular ejecta or from the NR simulation for the dynamical ejecta.
(-) is placed when the value does not affect the fit outcome.}
\centering
\begin{tabular}{l || c c | c || c c c c | c}
\multicolumn{9}{c}{Bolometric luminosity} \\
\hline
& \multicolumn{3}{c}{Secular wind} & \multicolumn{5}{c}{NR dynamical ejecta} \\
\hline
Model & $\kappa$ [${\rm cm^2~g^{-1}}$] & $T_\mathrm{floor}$ [K] & ${\rm err}_L$ & $\kappa_\mathrm{high~lat}$ [${\rm cm^2~g^{-1}}$] & $\kappa_\mathrm{low~lat}$ [${\rm cm^2~g^{-1}}$] & $T_\mathrm{Ni}$ [K] & $T_\mathrm{LA}$ [K] & ${\rm err}_L$ \\
\hline
\texttt{xkn-ts} & 0.7 & - & 0.32 & \cellcolor[HTML]{C0C0C0} 0.1 & \cellcolor[HTML]{C0C0C0} 5.0 & - & - & 0.32 \\
\texttt{xkn-ts} (fixed opacity) & 22.3* & - & 1.21 & * & * & - & - & 0.44 \\
\texttt{xkn-diff} & 4.9 & 154 & 0.12 & 1.0 & 10.7 & 5848 & 2150 & 0.12 \\
\texttt{xkn-diff} (fixed opacity) & 22.3* & 5999 & 0.31 & * & * & 4183 & 2112 & 0.14 \\
\hline
\multicolumn{9}{c}{} \\
\multicolumn{9}{c}{AB magnitudes} \\
\hline
& \multicolumn{3}{c}{Secular wind} & \multicolumn{5}{c}{NR dynamical ejecta} \\
\hline
Model & $\kappa$ [${\rm cm^2~g^{-1}}$] & $T_\mathrm{floor}$ [K] & ${\rm err}_m$ & $\kappa_\mathrm{high~lat}$ [${\rm cm^2~g^{-1}}$] & $\kappa_\mathrm{low~lat}$ [${\rm cm^2~g^{-1}}$] & $T_\mathrm{Ni}$ [K] & $T_\mathrm{LA}$ [K] & ${\rm err}_m$ \\
\hline
\texttt{xkn-ts} & 5.6 & 1026 & 1.49 & 1.8 & 6.0 & 1857 & 1019 & 0.91 \\
\texttt{xkn-ts} (fixed opacity) & 22.3* & 3242 & 3.25 & * & * & 3424 & 946 & 1.01 \\
\texttt{xkn-diff} & 19.0 & 797 & 1.06 & 0.2 & 49.4 & 1516 & 433 & 0.74 \\
\texttt{xkn-diff} (fixed opacity) & 22.3* & 984 & 1.08 & * & * & 1556 & 591 & 1.05 \\
\hline
\end{tabular}
\label{tab:fits}
\end{table*}

The results for all the different models and configurations considered are summarized in \reftab{tab:fits}.
As visible, the fit procedure returns reasonable fit parameters values falling in the prior intervals, with the exception of a minor number of cases useful to let the modelling limits emerge.
In general, both the fits on the bolometric luminosities and the magnitudes derived from the RT simulation show an overall improved fit quality when using \texttt{xkn-diff} with respect to the previous \texttt{xkn-ts} model.\\

\begin{figure*}
\includegraphics[width=\textwidth]{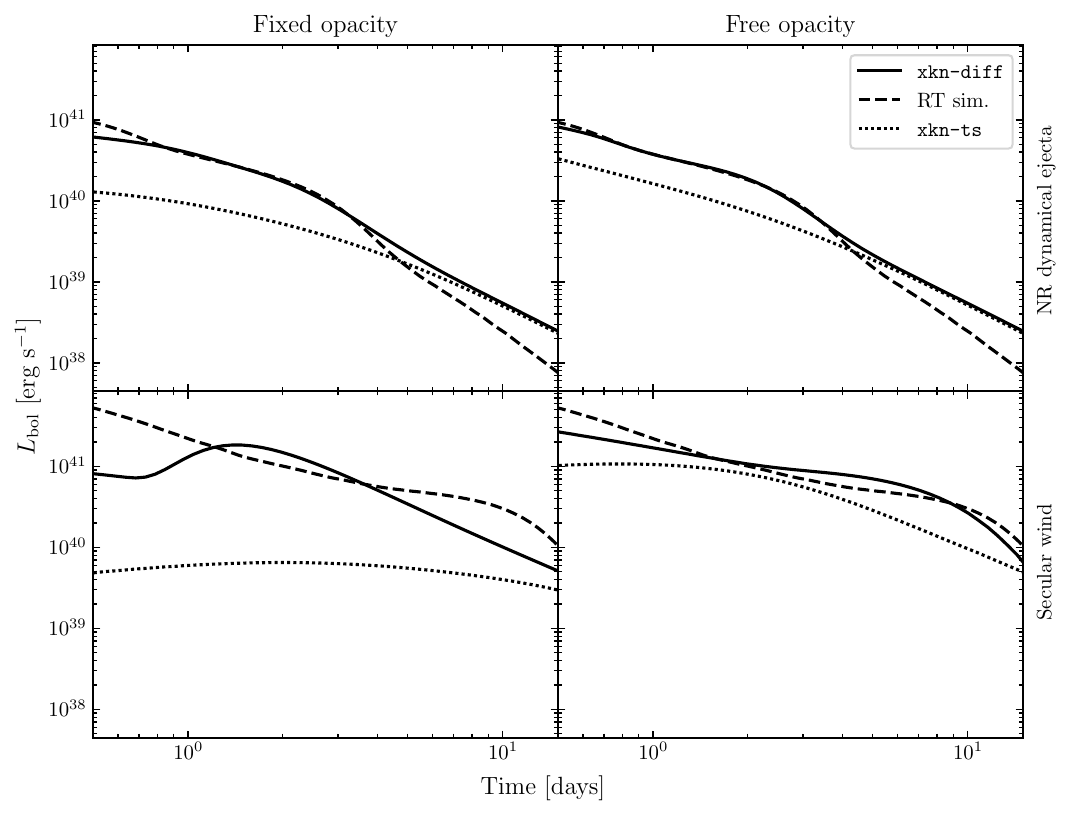}
\caption{Bolometric luminosity correspondent to the best fit parameters in the dynamical and the secular component configurations, as obtained by using \texttt{xkn-diff} and \texttt{xkn-ts} models, with and without a fixed opacity. Curves are shown in comparison to the data on which the fits are performed, derived from the RT calculations obtained with the code developed in \citet{Tanaka:2013ana,Kawaguchi:2018ptg,Kawaguchi:2020vbf}.}
\label{fig:lums_fit}
\end{figure*}

In particular, when fitting on the bolometric luminosity, \texttt{xkn-ts} is limited by having only the degree of freedom associated to the ejecta opacity (when the latter is left free to vary), since the temperature floor does not enter the luminosity calculation, as opposed to the \texttt{xkn-diff} case.
This difference arises because the floor temperature affects the late time photospheric radius in both models, but while in \texttt{xkn-ts} the latter is used only for the magnitudes computation through the Stefan-Boltzmann law and does not modify the volume of the radiative zone, in the \texttt{xkn-diff} case the photosphere position has a feedback on the allocation of mass to the optically thick and optically thin regimes, thus altering the bolometric luminosity as well.
As a result, for the case in which the opacity is prescribed using the value of the electron fraction, the bolometric luminosity in \texttt{xkn-ts} is completely fixed for both the secular and the dynamical component configurations, and the correspondent fit errors are the worst in the set.\\
On the other hand, in the \texttt{xkn-diff} model the floor temperature is adjusted in such a way to increase the late time agreement.
Indeed such parameter can ultimately affect light curves only when temperatures in the ejecta have decreased enough, as it is commonly assumed to be the case around a few days post-merger.
Specifically, in the secular component configuration this dependency drives the floor temperature to almost rail against the upper boundary, in order to accelerate the photosphere recession and maximize the amount of thin ejecta contributing to the late time emission.
One can note that in all cases, and more rapidly for the faster dynamical component, both semi-analytic models converge to the same curve, since in the \texttt{xkn-diff} model the treatment of the thin ejecta, which eventually constitutes the totality of the outflow, is analytically equivalent to the one used for the entire ejecta in \texttt{xkn-ts}.
Furthermore, since this treatment does not properly model the material opacity outside of the photosphere, varying the latter cannot affect the computed luminosity around $~10$ days, thus not improving the late time matching.\\
As visible in \reffig{fig:lums_fit}, in all the investigated configurations \texttt{xkn-ts} is sistematically underestimating the early time luminosity with respect to both the RT simulation and \texttt{xkn-diff} of a factor from a few to even multiple orders of magnitude depending on the specific case.
This evidence establishes qualitatively the error hierarchy in using an approximate scheme based on the calculation of the diffusion timescale of photons, versus an analytic solution of the simplified RT problem, with respect to a full RT simulation.
Therefore, once the opacity is left free to vary, the physiologic behaviour of \texttt{xkn-ts} is to compensate this systematic by lowering the latter to very small values in order to increase the emission brightness especially before $\sim1$ day, even incurring in the opacity boundaries in the dynamical component case.
Also the \texttt{xkn-diff} model is partially subject to the same mechanism, as visible specifically in the secular component case.
This result suggests a general limitation on using the bolometric luminosity to fit parameters related to local features of the ejecta configuration.
However, we also note that, especially for the dynamical component, the magnitude and shape of the bolometric luminosity are in good agreement with the ones derived from the RT calculations.

\begin{figure*}
\includegraphics[width=\textwidth]{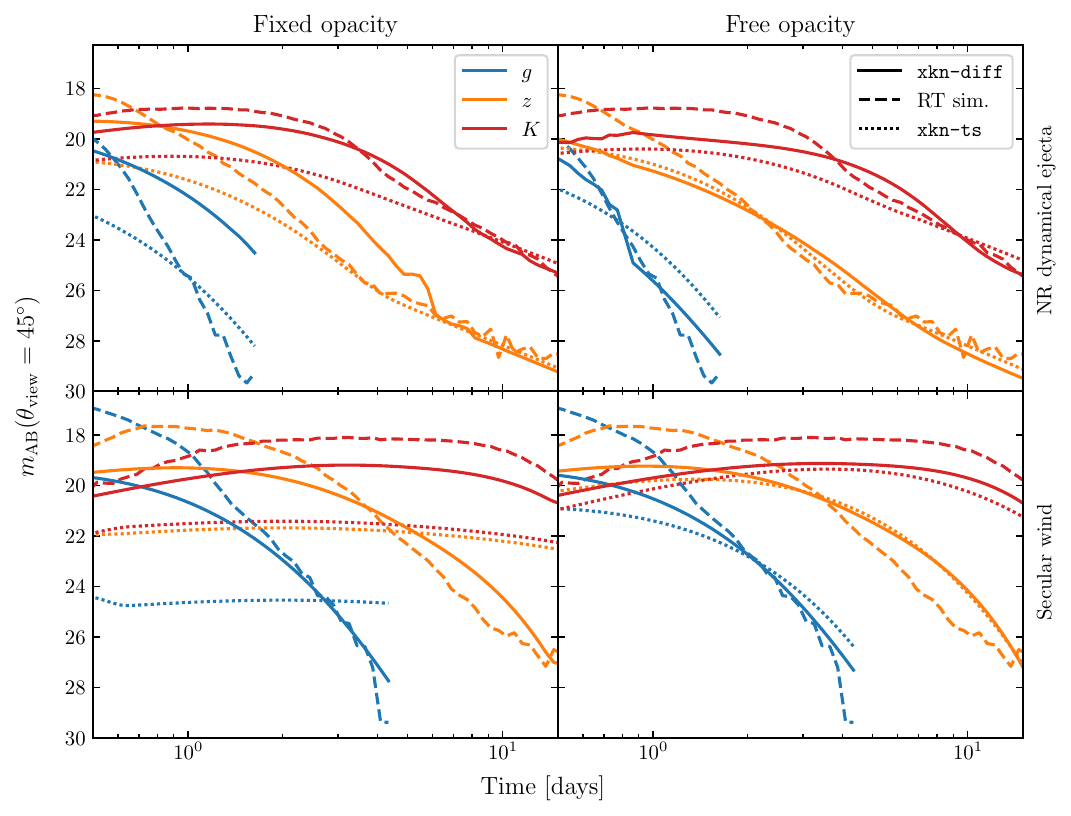}
\caption{AB magnitudes in the $g$, $z$ and $K$ filters as seen from a viewing angle $\theta_{\rm view}=45^\circ$, calculated for the best fit parameters in the dynamical and the secular component configurations, using \texttt{xkn-diff} and \texttt{xkn-ts} models, with and without a fixed opacity. Curves are shown in comparison to the data on which the fits were performed, derived from the RT calculations obtained with the code developed in \citet{Tanaka:2013ana,Kawaguchi:2018ptg,Kawaguchi:2020vbf}.}
\label{fig:mags_fit}
\end{figure*}

With respect to the fit on the bolometric luminosity, when the same procedure is applied to the AB magnitudes, the overall qualitative results remain roughly unaltered.
However, in such a case we include by construction more information, coming from different wavebands and viewing angles.
In addition, the temperature floor has a direct role in determining the color bands, since they strongly depend on the photosphere effective temperature, and thus this parameter influences the fit outcome regardless of the model employed.
Therefore, we obtain best fit values not necessarily close to the ones found in the previous case.\\
Both in the secular and the dynamical component configurations, we find the same error hierarchy as in the bolometric luminosity fits, with \texttt{xkn-ts} tipically underestimating the overall brightness up to a few days in all bands with respect to \texttt{xkn-diff}.
Furthermore, \reffig{fig:mags_fit} shows how magnitudes confirm the trend already evident for the bolometric luminosity, by which \texttt{xkn-diff} tends to underestimate the emission brightness at early times $\lesssim1$ day, indicating a model limitation.\\
The retrieved opacities are now generally higher for both models, with values which can also be closer to the ones fixed by the atomic calculations.
The fact that the opacity values differ significantly from the previous fits is not surprising, since in this case they are informed with the light curves in multiple filters as seen in edge-on and face-on configuration: especially in the dynamical ejecta configuration, the latter is valuable information in determining the opacity angular distribution with a better accuracy.
In addition, we recall that color bands in the model are derived by composing a sprectrum mainly based on pure black-body emission, which is therefore not able to reproduce the black-body deviations found in the RT calculations.
In particular, as pointed out in \citet{Gillanders:2022opm}, we note that realistically part of the UV radiation is reprocessed by the heavy elements into the optical and NIR bands, thus shifting the emitted energy distribution significantly, without an heavy alteration in the bolometric luminosity.
As a consequence, the more sensitive fits on the magnitudes retrieve opacity values which are increased in order to compensate for the lack of such feature in the model.
The floor temperatures derived from the magnitudes are substantially different from the ones derived from the bolometric luminosity.
This is due a combination of effects, whereby the floor temperature is not trivially connected to the final magnitudes and its value is subject to stronger variability.\\
On one side, higher floor temperatures are associated to stronger radiation fluxes at late times and, for a given energy emission rate, to smaller photospheric radii, with a net increase in the late broadband magnitudes.
Being this the only effect in \texttt{xkn-ts}, the magnitude fit finds the best temperature floor parameter value up to $\sim3400$ K, in order to compensate for the systematic underestimation of the model, partially relieving the opacity parameter from such burden.
This behaviour has also to be ascribed to our fit procedure, which tries to minimize quantitatively the separation between different curves, rather than trying to reproduce the same shape.
For this reason, the detailed values of floor temperature that we obtain are not meant to be reliable, but they can nevertheless highlight the internal structure of the model.\\
On the other side, on top of the above effect, as already pointed out for the bolometric luminosity fits, in the \texttt{xkn-diff} model higher floor temperatures also cause a faster decrease in the amount of optically thick ejecta at late times.
In particular, in this case the temperature floor recovery which results from such interplay cures the drift towards the upper boundary that is found in the bolometric luminosity fit for the secular ejecta configuration with fixed opacity.
As a general consequence, for \texttt{xkn-diff}, values are almost systematically and significantly lower than both \texttt{xkn-ts} and their counterparts in the previous fits, being in some cases down to only a few hundreds Kelvin degrees, and indicating a tendency to decrease the overall radiation fluxes in order to match color bands after a few days.\\

\section{Conclusions}\label{sec:conclusions}
In this work we presented the new framework \texttt{xkn} for the computation of the kilonova emission from compact binary mergers, starting from the characterization of the merger ejecta.
The framework allows for non-trivial ejecta structures as can be inferred from numerical relativity merger simulations, and it employs the results of recent efforts in nuclear astrophysics and atomic physics in terms of inputs for the kilonova model, such as radioactive heating rates from nuclear reaction network calculations \citep{Perego:2020evn} and grey opacities from systematic atomic structure calculations \citep{Tanaka:2019iqp}.
With respect to previous iterations, \texttt{xkn} includes the model \texttt{xkn-diff}, which encapsulate different possible semi-analytic solutions of the diffusion equation for the radiation energy density field, derived from the RT problem under the assumption of homologously expanding material \citep{Wollaeger:2017ahm}. \texttt{xkn-diff} constitutes an improvement with respect to previous semi-analytic models based on simpler laws of energy conservations and approximate radiation diffusion timescale estimations.
In addition, the model tracks the position of the ejecta photosphere in time in order to distinguish between the optically thick internal bulk and the optically thin external layers.
The latter are treated with a simplified shell model, which approximately accounts for the non-negligible contribution to the total luminosity coming from this region from a few days post-merger on.\\
We tested \texttt{xkn} models by comparing their results with the ones obtained from two-dimensional RT simulations obtained with the code developed in \citet{Tanaka:2013ana,Kawaguchi:2018ptg,Kawaguchi:2020vbf}, which are based on the same ejecta configurations.
In particular, we considered two representative scenarios, i.e. a lighter anisotropic dynamical component and a more massive spherical secular component, and we fit the free parameters of the model to the bolometric luminosity and the AB magnitudes in different filters, as seen from multiple viewing angles.
We found that \texttt{xkn-diff} is able to reproduce the overall behaviour of the light curves obtained from the RT simulations, with a better agreement with respect to the previous semi-analytic model, despite the simplified treatment of the decay energy thermalization process and of the ejecta opacity.
However, as highlighted by the fit procedure, the latter still constitutes a limitation to this modelling approach and it will be the subject of future improvements.
In particular, the average constant grey opacity values that the model employs are a crude approximation of the real effective opacity inside the ejecta, which significantly varies of more than one order of magnitude with time and across the different regions of the outflow, depending on the local temperature, density and composition.
As a result, the emission brightness at early times, i.e. around a few hours post-merger, predicted by the model in the fit procedure, can be systematically lower with respect to the RT calculation, of a factor of a few in the bolometric luminosity and of up to 2 magnitudes in the color bands.
We also note that the temperature floor, a secondary parameter in \texttt{xkn} which often appears in other semi-analytic models, is not easily constrained, since it is not trivially connected to the final magnitudes.\\
We conclude that \texttt{xkn} constitutes a valid tool to model the kilonova emission from compact binary mergers, with the main strength being its computational efficiency, which allows for extensive explorations of the ejecta parameter space in a reasonable time frame.
This is particularly useful in the context of the now thriving multi-messenger astronomy, whereas the kilonova is only one of the possible electromagnetic counterparts of the merger event.
Coupling this model with information from other sources, such as the GRB afterglow or the GW signal, in a statistical framework, can sinergically help to constrain the properties of the original binary, the central remnant or the merger ejecta, and thus to shed light on the nature of the detected event itself.

\section*{Acknowledgements}
GR acknowledges support by the Deutsche Forschungsgemeinschaft (DFG, German Research Foundation) – Project-ID 279384907 – SFB 1245.
GR acknowledges support by the State of Hesse within the Research Cluster ELEMENTS (Project ID 500/10.006).
SB acknowledges support from the Deutsche Forschungsgemeinschaft, DFG, Project MEMI number BE 6301/2-1.
SB acknowledges support by the EU Horizon under ERC Consolidator Grant, no. InspiReM-101043372 and from the Deutsche Forschungsgemeinschaft, DFG, Project MEMI number BE 6301/2-1.
KK acknowledges support by Grant-in-Aid for Scientific Research (JP20H00158, JP21K13912, JP23H04900) of JSPS/MEXT.
We thank Masaomi Tanaka for valuable discussions and feedback on the manuscript.
We thank Federico Schianchi for initial tests at the beginning of the project.

\section*{Data availability}
The kilonova framework presented in this work is publicly available at \url{https://github.com/GiacomoRicigliano/xkn}.

\bibliographystyle{mnras}
\bibliography{refs,local}

\bsp
\label{lastpage}
\end{document}